\documentclass[twocolumn,aps,superscriptaddress,floatfix]
{revtex4}

\setcitestyle{super}

\usepackage{amsmath}	
\usepackage{float}

\usepackage{ifpdf}
\ifpdf
	\usepackage[pdftex]{graphicx}
	\usepackage{epstopdf}
\else
	\usepackage[dvipdfm]{graphicx}
\fi

\usepackage{hyperref}
\hypersetup{
    pdftitle = 
{Quantum tele-amplification with a continuous variable superposition state},
    pdfauthor = 
{Jonas S. Neergaard-Nielsen, Yujiro Eto, Chang-Woo Lee, Hyunseok Jeong, 
 and Masahide Sasaki},
    pdfkeywords = {quantum optics, quantum information, quantum teleportation,
                   quantum key distribution, QKD, quantum communication,
                   coherent states, coherent state quantum computation},
    bookmarksopen=false
}

\newcommand{\ket}[1]{\left | #1 \right \rangle}
\newcommand{\bra}[1]{\left \langle #1 \right |}
\newcommand{\proj}[1]{\ket{#1} \bra{#1}}

\newcommand{\amp}[2]{\left \langle #1 | #2 \right \rangle}

\newcommand{\beq}{\begin{equation}}
\newcommand{\eeq}{\end{equation}}
\newcommand{\beqa}{\begin{eqnarray}}
\newcommand{\eeqa}{\end{eqnarray}}
\newcommand{\beqan}{\begin{eqnarray*}}
\newcommand{\eeqan}{\end{eqnarray*}}

\newcommand{\affA}{%
\affiliation{
     National Institute of Information and Communications Technology
     (NICT), \\
     4-2-1 Nukui-kitamachi, Koganei, Tokyo 184-8795, Japan}
     }
\newcommand{\affB}{%
\affiliation{
	Center for Macroscopic Quantum Control,
        Department of Physics and Astronomy, \\
        Seoul National University, Seoul, 151-747, Korea}
	}
\newcommand{\affC}{%
\affiliation{
	Department of Physics,
		Technical University of Denmark, 
		Fysikvej, 2800 Kgs.~Lyngby, Denmark}
	}
\newcommand{\affD}{%
\affiliation{
	Department of Physics, 
		Texas A\&M University at Qatar, 
		PO Box 23874, Doha, Qatar}
	}
\newcommand{\affE}{%
\affiliation{
	Department of Physics, 
		Gakushuin University, 
		1-5-1 Mejiro, Toshima-ku, Tokyo 171-8588, Japan}
	}
\newcommand{\affF}{%
\affiliation{
	Centre for Quantum Computation and Communication 
        Technology, School of Mathematics and Physics, 
        University of Queensland, Brisbane, Queensland 4072, 
        Australia}
	}

\begin{document}
\title{Quantum tele-amplification with
a continuous variable superposition state}
\date{\today}
\author{Jonas S. Neergaard-Nielsen}
\affA\affC%
\author{Yujiro Eto}
\affA\affE%
\author{Chang-Woo Lee}
\affB\affD%
\author{Hyunseok Jeong}
\affB\affF%
\author{Masahide Sasaki}%
\affA%
%




\begin{abstract}
Optical coherent states are classical light fields with high purity, 
and are essential carriers of information in optical networks. 
If these states could be controlled in the quantum regime, 
allowing for their quantum superposition 
(referred to as a Schr\"odinger cat state), 
then novel quantum-enhanced functions
such as coherent-state quantum computing (CSQC) 
\cite{Cochrane1999,Jeong2002,Ralph2003,Jeong2007,Lund2008},
quantum metrology 
\cite{Gerry2002,Joo2011}, 
and a quantum repeater \cite{Sangouard2010,Brask2010a}, 
could be realized in the networks.
Optical cat states are now routinely generated in the laboratories. 
An important next challenge is to use them 
for implementing the aforementioned functions. 
Here we demonstrate a basic CSQC protocol, 
where a cat state is used as an entanglement resource 
for teleporting a coherent state with an amplitude gain. 
We also show how this can be extended to 
a loss-tolerant quantum relay of 
multi-ary phase-shift keyed coherent states. 
These protocols could be useful both 
in optical and quantum communications.
\end{abstract}

\maketitle

%
%

Among various optical implementations of quantum information
processing (QIP), coherent-state quantum computing (CSQC) is of special
interest for enhancing the performance of optical communications, 
where information is encoded into coherent states.
These are the only states that can be transmitted
preserving the state purity even through a lossy channel 
since they are eigenstates of the annihilation operator, 
$\hat a \vert\alpha\rangle=\alpha\vert\alpha\rangle$. 
Hence, a simple classical encoding with coherent states 
can be the optimal strategy of the transmitter 
to achieve the ultimate capacity of a lossy optical channel 
\cite{Giovannetti2004}. 
On the receiver side, 
the sequence of coherent-state pulses should be decoded 
fully quantum mechanically 
by employing a collective measurement with CSQC 
\cite{Sasaki1998}.
This scheme can realize communication
with larger capacity, 
beating the conventional homodyne limit of optical communications 
\cite{Waseda2010}.
Although practical implementation of CSQC remains a big challenge,
recent progress in generating 
\cite{Ourjoumtsev2006,Neergaard-Nielsen2006, Wakui2007, Ourjoumtsev2007a, Takahashi2008} 
and
manipulating 
\cite{Neergaard-Nielsen2010, Lee2011} 
optical cat states makes it realistic to implement its basic
building blocks. In this paper, we propose and demonstrate the
first operational application of cat states 
for QIP, where
a cat state is used as the entanglement resource for
teleporting a coherent state with an amplitude gain. 
We also propose its new application to quantum key distribution (QKD),
namely a loss-tolerant quantum relay of multi-ary
phase-shift keyed (M-PSK) coherent states that does not assume
a trusted node.
We present its proof-of-principle
demonstration with binary PSK states.

%
%

\vspace{1em}


\begin{figure}
\begin{center}
\includegraphics[width=0.90\linewidth]
{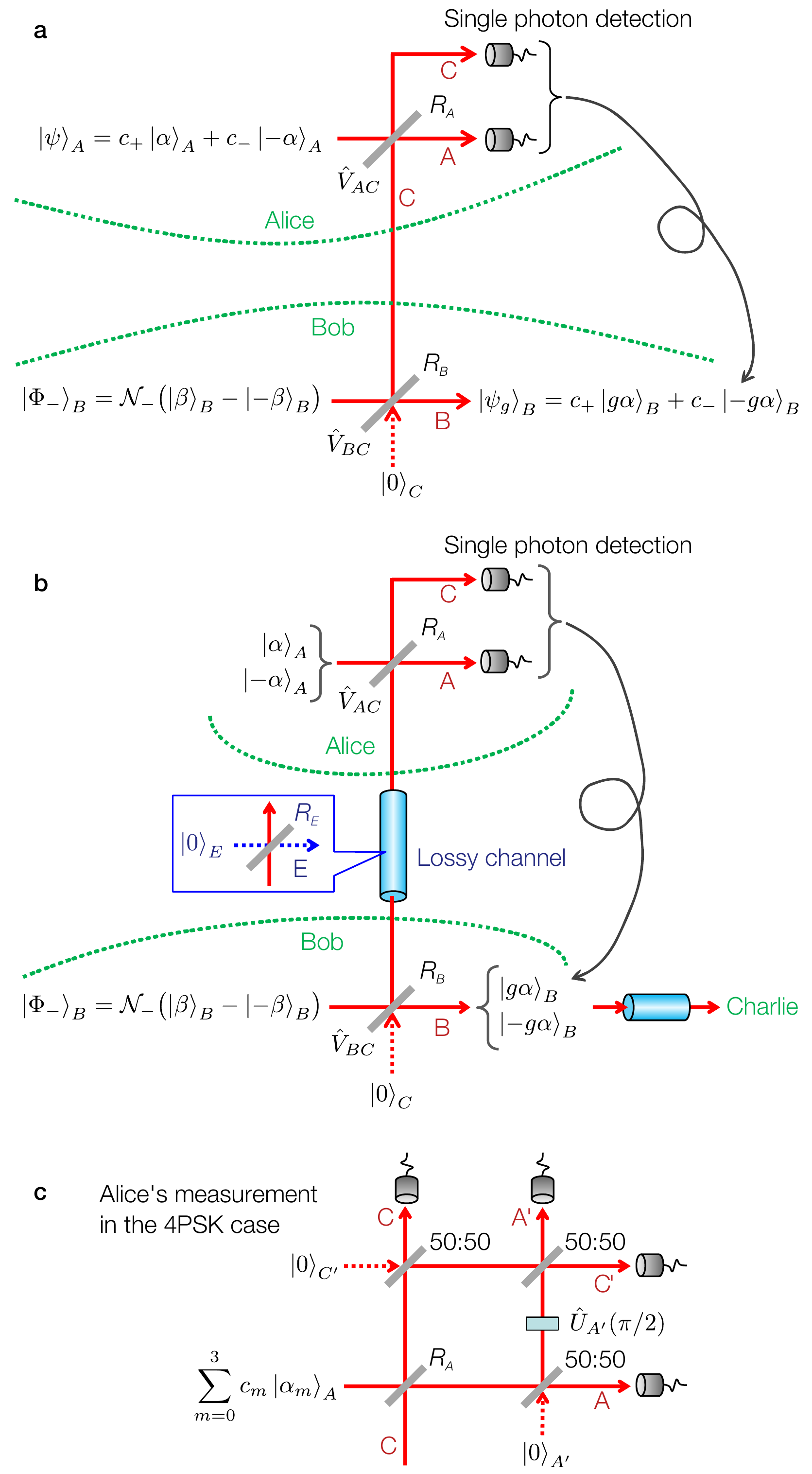}   %
\caption{\label{Tele_Amp_Scheme_BPSK_4PSK}
Scheme of quantum tele-amplification and quantum relay. 
{\bf a} Tele-amplification of binary cat-state 
in an ideal lossless channel. 
$R_A$ and $R_B$ are the reflectivities of the BSs. 
{\bf b} 
Loss tolerant quantum relay. 
$R_E$ is the reflectivity of the BS 
which models the lossy channel. 
{\bf c} 
Alice's four-port measurement for the case of 
4PSK states. 
}
\end{center}
\end{figure}

The basic scheme of teleportation from Alice to Bob of 
a cat state qubit
$\ket{\psi}_A=c_+ \ket{\alpha}_A+c_- \ket{-\alpha}_A$, 
which is a variation of the schemes in refs.
\citenum{vanEnk2001} and \citenum{Jeong2001},
is depicted in Fig. 
\ref{Tele_Amp_Scheme_BPSK_4PSK}{\bf a}. 
Bob prepares an odd cat state 
$
\ket{\Phi_-(\beta)}_B
=
{\cal{N_-}}\bigl(\ket{\beta}_B -\ket{-\beta}_B \bigr) 
$ 
with normalization $\cal{N_-}$ and splits it into an entangled cat state
over paths B and C via a beam-splitter (BS)
$\hat V_{BC}$  with reflectivity $R_B$.
He sends one part of it to Alice at port C. 
She then combines it on an $R_A$-reflectivity BS 
with her input $\ket{\psi}_A$ at port A as 
\begin{equation}
\label{after beamsplitters}
\ket{\Psi}_{ABC}
=
\hat V_{AC} \ket{\psi}_A
\hat V_{BC} \ket{\Phi}_B \ket{0}_C. 
\end{equation}
She finally measures modes A and C by single-photon detectors. 
By conditioning port B on her measurement result, 
Bob can restore Alice's input.

The amplitude of the resource cat state is set as 
$
\beta = \alpha \sqrt{ (1-R_A)/R_A R_B }, 
$
such that the components at port A turn into 
either the vacuum or a non-vacuum state. 
Then, when Alice's detectors register a single photon 
at port A and nothing at port C 
-- denoted (1,0) -- 
Bob unambiguously obtains the state (see Appendix \ref{appA}) 
\begin{equation}
\label{output in ideal tele-amplification:binary}
{}_{AC}\langle1,0\vert\Psi\rangle_{ABC}
\propto
 c_+ \ket{-g\alpha }_B
+c_- \ket{ g\alpha }_B,
\end{equation}
where  
$
g=\sqrt{ (1-R_A) (1-R_B) / R_A R_B  }
$
is the gain parameter. 
By a simple $\pi$-phase shift, 
it can be transformed to Alice's input $\ket{\psi}_A$,
but with a modified amplitude $\alpha' = g\alpha$.
This process, previously suggested in ref. \citenum{Ralph2003},
we will call \textsl{tele-amplification}.

Unfortunately this tele-amplification is vulnerable to losses.
Suppose the channel between Alice and Bob is subject 
to a linear loss $R_E$. The amplitude of the resource cat state
should then be chosen as
\begin{equation}
\label{amplitude of resource cat}
\beta =\sqrt{ \frac{1-R_A}{R_A R_B (1-R_E)} } \alpha .
\end{equation}
After conditioning on Alice's detection event (1,0),
Bob's state gets entangled with an external mode E as
\begin{equation}
\label{lossy tele-amplification}
\ket{\psi}_A \ket{0}_E
\mapsto
 c_+
\ket{-g\alpha }_B
\ket{ \varepsilon }_E
+
c_-
\ket{ g\alpha }_B
\ket{-\varepsilon }_E
\end{equation}
with the modified gain including the loss rate $R_E$ 
\begin{equation}
\label{gain for lossy tele-amplification}
g=\sqrt{ \frac{ (1-R_A) (1-R_B) }{ R_A R_B (1-R_E) }  }.
\end{equation}
Here 
$\varepsilon=\sqrt{(1-R_A)R_E/R_A(1-R_E)}\alpha$. 
Thus, the output at Bob is
generally in a decohered state.

%
%

One can, however, 
see that if Alice's inputs are restricted to classical components, 
$\ket{\alpha}$ or $\ket{-\alpha}$, as in Fig.
\ref{Tele_Amp_Scheme_BPSK_4PSK}{\bf b}, 
the output state can be completely disentangled 
from the external mode. 
This means that the coherent states can be tele-amplified faithfully 
to the target states even through the lossy channel as
\begin{equation}
\label{BPSK tele-amplification}
\ket{\pm\alpha}_A \mapsto \ket{\pm g\alpha}_B. 
\end{equation}
This is referred to as \emph{loss-tolerant quantum relay}. 
In this context, 
Bob plays the role of an intermediate node, 
restores the target states $\ket{\pm g\alpha}_B$, 
and sends them to the terminal node, Charlie.

This simplest binary case can be extended into M-PSK coherent states. 
Let us show it for the 4-PSK case,
$\ket{\alpha_m},\,(\alpha_m=i^m\alpha,\,m=0,1,2,3)$. 
Bob should prepare a 4-component cat state
\begin{equation}
\ket{\Phi}_B
=
{\cal N} \displaystyle\sum_{k=0}^3 i^k \ket{i^k\beta}_B
\end{equation}
as a resource. 
This state is beam-split, and is shared with Alice.
We set $R_A=0.5$.
As in Fig. \ref{Tele_Amp_Scheme_BPSK_4PSK}{\bf c}, 
Alice performs a four-port single-photon detection
at paths A, A', C, and C' on this state.
Depending on the set of results at the four ports, 
(A, A', C, C'),
the inputs are tele-amplified as
\begin{equation}
\label{QPSK tele-amplification}
\begin{array}{llll}
\ket{\alpha_m} &\mapsto& \ket{g\alpha_m},
    \,\,&\mathrm{for}\,\, (0,1,1,1),
\\
\ket{\alpha_m} &\mapsto& \ket{ig\alpha_m},
    \,\,&\mathrm{for}\,\, (1,0,1,1),
\\
\ket{\alpha_m} &\mapsto& \ket{-g\alpha_m},
    \,\,&\mathrm{for}\,\, (1,1,0,1),
\\
\ket{\alpha_m} &\mapsto& \ket{-ig\alpha_m},
    \,\,&\mathrm{for}\,\, (1,1,1,0).
\end{array}
\end{equation}
Thus, the simple tele-amplification is performed for
the result (0,1,1,1).
Moreover the output state can be switched 
to another element  
by choosing an appropriate click pattern at Alice
(see Appendix \ref{appB}).

\begin{figure*}
\hspace{10mm}
\begin{center}
\includegraphics[width=0.85\linewidth]
{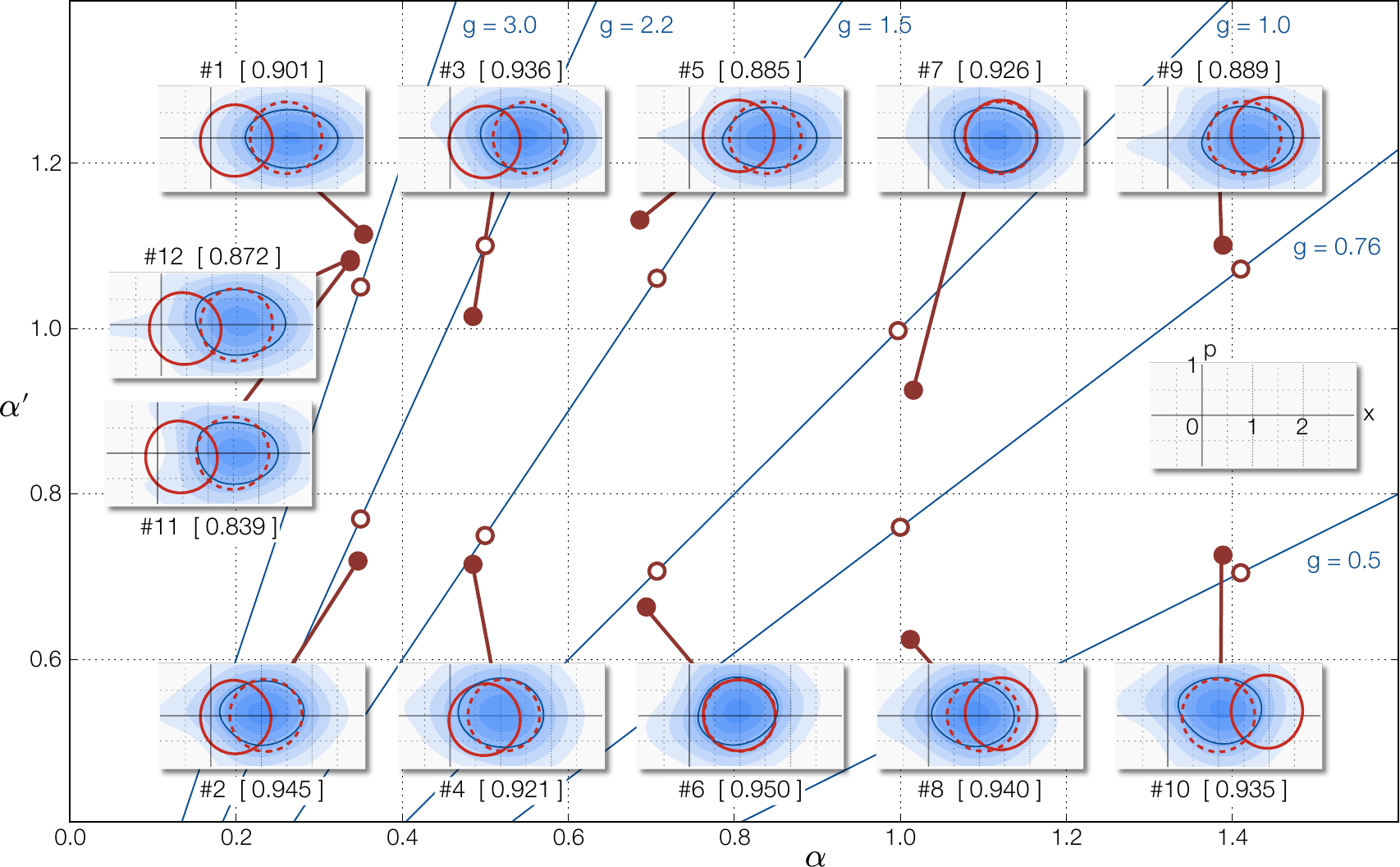}%
\caption{\label{Experimental_result}
Measured results for the twelve cases. 
The blue straight lines represent $\alpha'=g\alpha$. 
The open circles represent the sets of $\alpha$ and $g_\mathrm{tg}$ 
in Table I. 
The fidelities between the measured states and the targeted
$\ket{g_\mathrm{tg}\alpha}$ 
are indicated by the numbers in brackets.
Filled circles represent the amplitudes $\alpha$ ($\alpha'$) 
of the coherent states that have the maximum fidelity 
with the measured input (output) states. 
\textbf{Insets:} 
Output state Wigner functions are shown in blue contour plots. 
The red solid and dashed circles are 
for the input and ideal targeted states, respectively.
}
\end{center}
\end{figure*}

The faithful relay itself can also be realized in a classical way, 
where 
Bob at the intermediate node 
performs an unambiguous state discrimination 
on the signal state, 
reproduces an amplified state for his confident result, 
and finally resends it to Charlie. 
The success probabilities
of the two methods are compared in Appendix \ref{appC}.
This classical relay cannot, however, be applied to 
a QKD relay node without the trusted node assumption. 
In contrast, a quantum relay can be carried out 
in the fully quantum domain, 
without Bob's knowing the signal state itself, 
though at the expense of preparing the entangled cat state, 
and an appropriate entanglement verification session. 
Similar ideas  
for single-photon QKD were presented in refs.
\citenum{Jacobs2002} and \citenum{Collins2005}.

Our loss-tolerant quantum relay is particularly useful 
for extending the distance of QKD which uses PSK coherent states, 
such as B92 and BB84 \cite{Koashi2004,Tamaki2009,Lo2007}. 
Although the secure key generation probabilities 
at short distances slightly degrades from the original PSK-BB84, 
they can remain at reasonable levels up to much longer distances 
by the loss-tolerant quantum relay (see Appendix \ref{appG}).

%
%
\vspace{1em}

We carried out an experimental demonstration of the tele-amplification 
in the simplest case of binary PSK as in
Fig.~\ref{Tele_Amp_Scheme_BPSK_4PSK}{\bf b} 
to realize Eq. \eqref{BPSK tele-amplification}. 
The resource cat state $\ket{\Phi_-}$ was generated 
by photon-subtraction from a squeezed vacuum 
with anti-squeezing along the real axis in phase space (Methods). 
Bob's BS was set to $R_B=0.1$. 
For a given desired gain $g$, 
we varied $R_A$ according to Eq. \eqref{gain for lossy tele-amplification}. 
The resource cat-state amplitude $\beta$,
experimentally tuned by the squeezing level, was then set 
by Eq.~\eqref{amplitude of resource cat}. 
The detector at port C was omitted
with negligible effect on the outcomes since
the events (A,C)=(1,1) would be rare.
Bob's output state was characterized by homodyne tomography.

We tested twelve settings as summarized in Table I. 
The five different input amplitudes $\alpha$ 
were real and ranged between 0.35 and 1.4. 
The protocol was carried out only for $\ket{\alpha}$ 
because the outcome for $\ket{-\alpha}$ 
would be trivially identical. 
The measured results are shown in 
Fig.~\ref{Experimental_result}. 
The blue straight lines are gain curves 
$\alpha'=g\alpha$ in the $(\alpha,\alpha')$ diagram. 
The open circles plotted along these lines
represent sets of $\alpha$ and $g_\mathrm{tg}$ 
in Table I. 
The Wigner functions of the tomographically reconstructed
tele-amplified output states $\hat\rho_\mathrm{out}$ are shown as
blue contour plots in the insets.
One contour level is highlighted
for comparison with 
the targeted states $\ket{g_\mathrm{tg}\alpha}$ (red dashed) 
and 
the actual input states 
$\hat\rho_\mathrm{in} \approx \ket{\alpha}\!\bra{\alpha}$ 
(red solid, also characterized by homodyne tomography). 
The discrepancies between $\hat\rho_\mathrm{out}$
and $\ket{g_\mathrm{tg}\alpha}$ are due to imperfections, 
including the deviation of the photon-subtracted state 
from the ideal resource cat, losses, impurity, 
and Alice's use of an on/off detector 
instead of two single-photon detectors (see Appendix \ref{appC}). 
For each setting, we calculate which coherent states
$\ket{\alpha}, \ket{\alpha'}$ have the highest fidelity with the
measured input and output states, respectively, that is,
$\alpha=\mathrm{argmax}_\gamma
        \langle\gamma|\hat\rho_\mathrm{in}|\gamma\rangle$ 
and 
$\alpha'=\mathrm{argmax}_{\gamma'}
        \langle\gamma'|\hat\rho_\mathrm{out}|\gamma'\rangle$.
These $(\alpha, \alpha')$ pairs are marked as filled circles.

Despite the imperfections, the tele-amplification succeeded 
with high fidelities 
$\mathcal{F} = \langle g_\mathrm{tg}\alpha|\hat\rho_\mathrm{out}| g_\mathrm{tg}\alpha\rangle$
between 0.89 and 0.95 as shown next to each inset. 
The obtained amplitudes (filled circles) 
are close to the targeted ones (open circles) 
in almost all cases. 
The success probabilities were in the range 0.3\% to 0.65\% (Methods). 
For larger $\alpha'$, the Wigner function shapes are 
slightly elongated 
due to the larger squeezing needed to produce those states. 
We note that 
our experimental settings were not fully optimized 
by taking into account spectral mode mismatch 
between the resource cat and input coherent states 
as well as between the APD and homodyne detectors. 
Had we done it, 
we estimate the achieved fidelities to have been 0.94--0.99.

The settings \#11 and 12 had an additional 
80\% loss ($R_E=0.8$) in the channel from Bob to Alice. 
In \#11, $R_A=0.5$ for the original lossless setting (as in \#1), 
while 
in \#12, $R_A=0.83$ as optimized according to 
Eq. \eqref{gain for lossy tele-amplification}, 
resulting in success probabilities of 0.17\% and 0.11\%.
The fidelities with the target state are
as high as 0.839 and 0.872, respectively, 
as compared with 0.901 in the lossless case. 
This demonstrates the loss tolerance of the protocol.

\setlength{\tabcolsep}{10pt}

\begin{table}
    \begin{tabular}{c|lllll|l}
        \# & $\alpha$ & $g_{\mathrm{tg}}$ & $\beta$ & $R_A$
                      & $R_E$ & $\mathcal{F}$ \\ \hline
        1  & 0.35 &  3.0 & 1.11 & 0.50 & 0 & 0.901 \\ 
        2  & 0.35 &  2.2 & 0.81 & 0.65 & 0 & 0.945 \\ 
        3  & 0.50 &  2.2 & 1.16 & 0.65 & 0 & 0.936 \\ 
        4  & 0.50 &  1.5 & 0.79 & 0.80 & 0 & 0.921 \\ 
        5  & 0.71 &  1.5 & 1.12 & 0.80 & 0 & 0.885 \\ 
        6  & 0.71 &  1.0 & 0.75 & 0.90 & 0 & 0.950 \\ 
        7  & 1.00 &  1.0 & 1.05 & 0.90 & 0 & 0.926 \\ 
        8  & 1.00 & 0.76 & 0.80 & 0.94 & 0 & 0.940 \\
        9  & 1.41 & 0.76 & 1.13 & 0.94 & 0 & 0.889 \\ 
        10 & 1.41 & 0.50 & 0.74 & 0.97 & 0 & 0.935 \\ 
        \hline
        11 & 0.35 &  3.0 & 1.11 & 0.50 & 0.8 & 0.839 \\
        12 & 0.35 &  3.0 & 1.11 & 0.83 & 0.8 & 0.872 \\
    \end{tabular}
\caption{Desired tele-amplification for the twelve settings of 
input coherent states and gains. 
$g_\mathrm{tg}$ is the targeted gain.
The last column shows the obtained teleportation fidelities.} 
\label{tbl:experimental_realizations}
\end{table}

%
%
\vspace{1em}

Teleportation of a cat-state qubit as in 
Eqs.
(\ref{after beamsplitters}-
\ref{output in ideal tele-amplification:binary})
is a prerequisite for CSQC. 
Interestingly, the tele-amplification allows
to convert between different amplitude qubit bases.
Although we previously generated such arbitrary cat qubits
\cite{Neergaard-Nielsen2010, Takeoka2010}, 
it was not feasible to tele-amplify them with the current setup
since three simultaneous APD clicks would be needed. 
Instead we simulated this protocol 
by accurately modelling the current experiment 
including all relevant practical imperfections 
(see Appendix \ref{appF}). 
Figure \ref{qubittele} shows the average fidelities 
between the teleported state for an input cat-state qubit 
and an output state from the model (Methods). 
For a wide range of input amplitudes $\alpha$ and 
output amplitudes $\alpha'$,
it is possible to surpass the classical limit of 2/3.

\begin{figure}
\begin{center}
\includegraphics[width=.95\linewidth]
{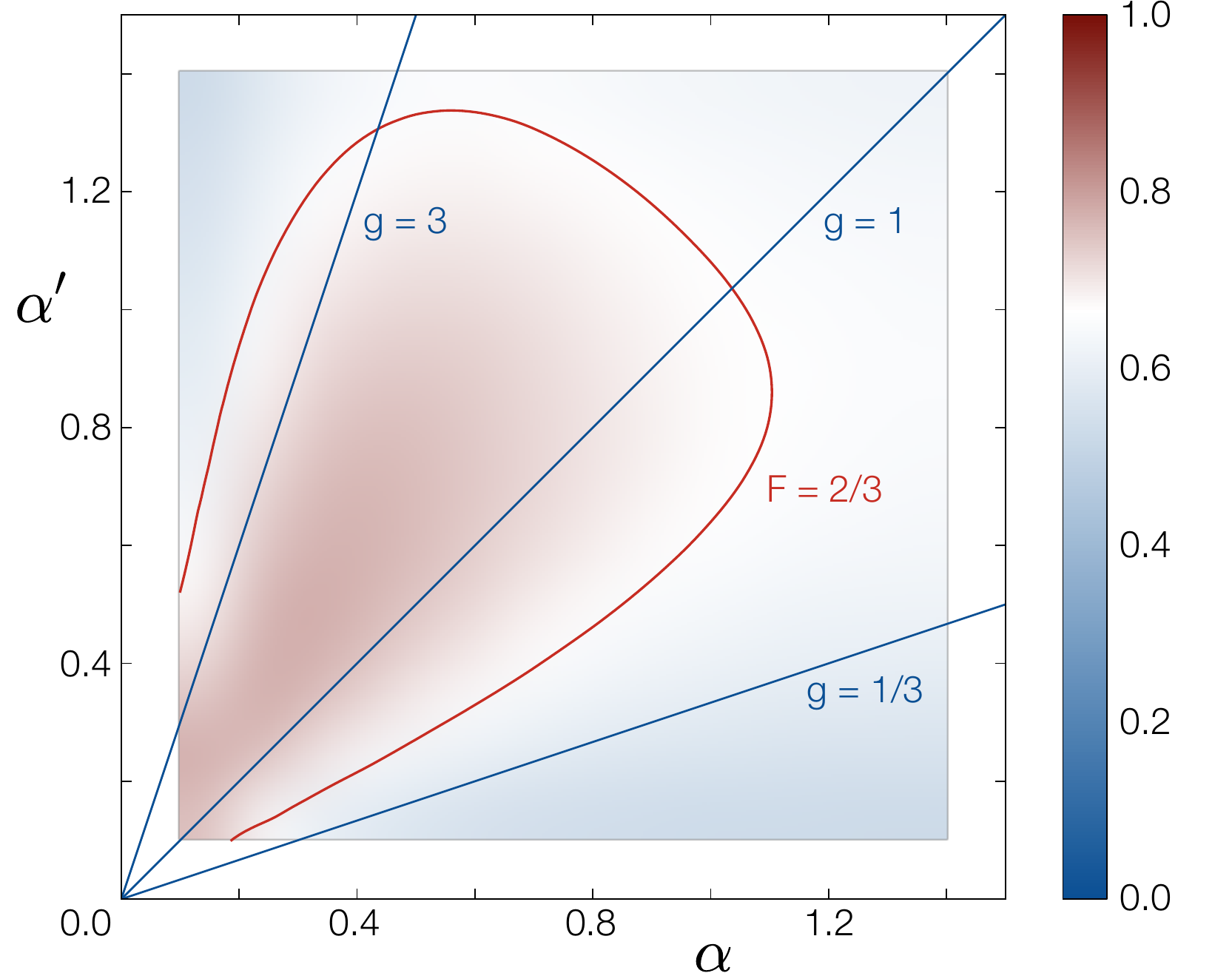}   %
\caption{\label{qubittele}
Simulated average qubit teleportation fidelity as a function of 
input ($\alpha$) and output state ($\alpha'$) amplitudes. 
All relevant practical imperfections in our experimental setup,
as described in Appendix \ref{appF}, 
are taken into account. 
The red curve labelled ``F = 2/3''
indicates the classical teleportation bound.
}
\end{center}
\end{figure}

%
%
\vspace{1em}

Finally we make a brief comparison between our scheme and 
the quantum noiseless amplifier with single-photon ancilla 
\cite{Xiang2010,Zavatta2010,Ferreyrol2011}. 
The latter is intended to noiselessly amplify 
a coherent state with an unknown amplitude 
at the cost of the success probability. 
In contrast, 
our scheme assumes the known amplitude $\alpha$ 
but instead enables one to tele-amplify PSK coherent states 
over a lossy channel 
with perfect fidelity and high success probability. 
It can also implement, in principle, 
the teleportation of their arbitrary superpositions.

In summary, we presented tele-amplification and loss-tolerant quantum relay 
of coherent states 
as the first operational application of optical cat states. 
The scheme is an essential building block for CSQC
as well as quantum communications.

\section*{Methods}


\noindent
{\bf Experiment} 
We generated the squeezed vacua at 860 nm wavelength from an
OPO (optical parametric oscillator) continuously pumped with 
pump parameters between 0.15 and 0.31, corresponding to $\beta$ values
of 0.78 to 1.15.
We tapped off 5\% of the squeezed beam on a BS 
and guided it to an APD. 
A click of the APD heralded the subtraction of a photon from the main beam
\cite{Wakui2007, Takeoka2010}.
The state thus generated is a close approximation to 
the odd cat state $\ket{\Phi_-}$, 
and has been shown to provide near-perfect teleportation performance 
\cite{Branczyk2008}.


Whenever Alice's APD clicked simultaneously with the heralding signal of
the single-photon subtraction for the resource cat-state generation, the
tele-amplification was successful, and we recorded a trace of the
homodyne signal of Bob's output state. 
The success probability is given by the ratio 
of the simultaneous click rate ($\sim$3--28 $s^{-1}$) 
to 
the photon subtraction click rate ($\sim$1000--4500 $s^{-1}$).
It is mainly limited by detector and spectral filtering efficiency.
To build the homodyne tomogram,
we repeated this procedure 6000--24000 times for each fixed input state,
with the local oscillator of the homodyne detector locked at phases
$-180^{\circ}, -150^{\circ}, \ldots, 150^{\circ}$ with respect to the
input state. Note
that the protocol succeeds as a single shot for an unknown input state
-- the repeated measurements with identical inputs are only needed for
characterizing the process by homodyne tomography.

Alice's input states were independently characterized 
by homodyne tomography at port C 
by setting $R_A=1$. 
To determine the input states accurately just at Alice's BS, 
we correct their reconstruction for the detection efficiency 
and the propagation losses from that point
to the homodyne detector. This total efficiency amounts to 88\%.
Likewise, in the reconstruction of Bob's output states we correct for
the overall detection efficiency of 94\% but not for any propagation
losses. 

A more detailed description of the experimental setup and the state
characterization can be found in Appendix \ref{appE}.

\noindent
{\bf Simulation of cat-state qubit teleportation} 
A cat-state qubit can be represented on a Bloch sphere as
\begin{eqnarray}
\ket{\psi(\alpha,\theta,\phi)}
&=&c_+ \ket{\mathrm{\alpha}}
  +c_- \ket{\mathrm{-\alpha}}
\nonumber\\
&=& \mathrm{cos}\frac{\theta}{2} \ket{\Phi_+(\alpha)}
+e^{ i \phi} 
   \mathrm{sin}\frac{\theta}{2} \ket{\Phi_-(\alpha)},
\nonumber
\end{eqnarray}
where 
$
|\Phi_\pm(\alpha)\rangle=
{\cal{N}}_\pm (\ket{\alpha} \pm \ket{-\alpha}) 
$ are the even/odd cat states with 
${\cal{N}}_\pm= 1/\sqrt{2(1\pm e^{-2\alpha^2})}$ 
and
$c_\pm =
  {\cal{N}}_+ \mathrm{cos}\frac{\theta}{2}
  \pm {\cal{N}}_- e^{ i \phi} \mathrm{sin}\frac{\theta}{2} .
$

Given an input state $\ket{\psi(\alpha,\theta,\phi)}$, 
our model of the experiment, described in Appendix \ref{appF},
returns a teleported output state
$\hat{\rho}_{\alpha,\theta,\phi}$.
To quantify the performance of the qubit teleportation
for specific settings of $\alpha$ and 
$\alpha'=g_\mathrm{tg}\alpha$, 
we calculate the average fidelity of the teleported state 
with the target state by integrating over the Bloch sphere:
\[
\mathcal{F}^{\mathrm{avg}}_{\alpha\rightarrow\alpha'} = 
\int d\phi d\theta \frac{\sin\theta}{4\pi} 
\bra{\psi(\alpha',\theta,\phi)} 
\hat{\rho}_{\alpha,\theta,\phi}
\ket{\psi(\alpha',\theta,\phi)}.
\]
The results for a range of amplitude settings are plotted in
Fig. \ref{qubittele}.


\paragraph*{Acknowledgments} We acknowledge helpful discussions
with K. Wakui, M. Takeoka, K. Hayasaka, 
M. Fujiwara, T.~C. Ralph, A.~P. Lund, K. Tamaki, and
M. Koashi.
This work was partly supported by the
Quantum Information Processing Project in 
the Program for World-Leading Innovation Research 
and Development on Science and Technology (FIRST)
and 
by a National Research Foundation of Korea (NRF) grant 
funded by the Korean Government 
(Ministry of Education, Science, and Technology) (No. 2010-0018295).

\paragraph*{Author contributions}
M.S. and J.S.N-N. formulated the basic protocol of tele-amplification 
and loss-tolerant quantum relay, 
inspired by a teleportation scheme by C.-W.L. and H.J..
J.S.N-N. and Y.E. carried out the experiment. 
J.S.N-N., C.-W.L., M.S. and H.J. 
performed the theoretical calculations. 
J.S.N-N. and M.S. wrote the paper with discussions
and input from all the authors.


\onecolumngrid\newpage\twocolumngrid
\newpage

\appendix

%
\section{Tele-amplification of a binary component cat-state}
\label{appA}
%

We first describe the tele-amplification of 
a binary component cat-state 
$
\ket{\psi(\alpha)}_A
=
c_0 \ket{\alpha}_A
+c_1 \ket{-\alpha}_A
$ 
through a lossless channel. 
The whole three-mode state after the beam-splitting operation 
in Fig. 1{\bf a} is given by 
\begin{widetext}
\beqa
\ket{\Psi}_{ABC}
&=&
\hat V_{AC} \ket{\psi(\alpha)}_A
\hat V_{BC} \ket{\Phi}_B \ket{0}_C
\nonumber\\
&=&
{\cal{N}}c_0
\ket{ \sqrt{1-R_A} \alpha - \sqrt{R_A R_B}\beta }_A
\ket{ \sqrt{1-R_B} \beta }_B
\ket{-\sqrt{R_A} \alpha - \sqrt{(1-R_A) R_B}\beta }_C 
\nonumber\\
&-&
{\cal{N}}c_0
\ket{ \sqrt{1-R_A} \alpha + \sqrt{R_A R_B}\beta }_A
\ket{-\sqrt{1-R_B} \beta }_B
\ket{-\sqrt{R_A} \alpha + \sqrt{(1-R_A) R_B}\beta }_C
\nonumber\\
&+&
{\cal{N}}c_1
\ket{-\sqrt{1-R_A} \alpha - \sqrt{R_A R_B}\beta }_A
\ket{ \sqrt{1-R_B} \beta }_B
\ket{ \sqrt{R_A} \alpha - \sqrt{(1-R_A) R_B}\beta }_C
\nonumber\\
&-&
{\cal{N}}c_1
\ket{-\sqrt{1-R_A} \alpha + \sqrt{R_A R_B}\beta }_A
\ket{-\sqrt{1-R_B} \beta }_B
\ket{ \sqrt{R_A} \alpha + \sqrt{(1-R_A) R_B}\beta }_C
\eeqa
\end{widetext}
where 
${\cal{N}}=1/\sqrt{2[1 - \exp(-2|\beta|^2)]}$ 
is the normalization of the resource cat state $\ket{\Phi}_B$. 
We now impose a condition
on the amplitude of the resource cat state
\beq\label{amplitude of resource cat:binary}
\beta =\sqrt{ \frac{1-R_A}{R_A R_B} } \alpha
\eeq
such that
the components at port A turn into 
either of the vacuum or non-vacuum states 
as
\beqa\label{Psi_ABC_after_nulling}
\ket{\Psi}_{ABC}
&=&
{\cal{N}}c_0
\ket{ 0 }_A
\ket{ g\alpha }_B
\ket{-\frac{1}{\sqrt{R_A}}  \alpha }_C
\nonumber\\
&-&
{\cal{N}}c_0
\ket{ 2\sqrt{1-R_A} \alpha }_A
\ket{-g\alpha }_B
\ket{ \frac{1-2R_A}{\sqrt{R_A} } \alpha }_C
\nonumber\\
&+&
{\cal{N}}c_1
\ket{-2\sqrt{1-R_A} \alpha }_A
\ket{ g\alpha }_B
\ket{-\frac{1-2R_A}{\sqrt{R_A} } \alpha }_C
\nonumber\\
&-&
{\cal{N}}c_1
\ket{ 0 }_A
\ket{-g\alpha }_B
\ket{ \frac{1}{\sqrt{R_A}}  \alpha }_C
\eeqa
with the gain  
\beq\label{gain in the lossless case}
g=\sqrt{ (1-R_A) (1-R_B) / R_A R_B  } .
\eeq

Alice then performs single photon detection on paths A and C 
as shown in 
Fig. 1{\bf a}, 
and selects single photon at port A and nothing at port C --
denoted (1,0). 
Then Bob can unambiguously exclude the first and fourth terms 
in Eq. (\ref{Psi_ABC_after_nulling}), and 
has the state
\beq\label{output in ideal tele-amplification:binary}
{}_{AC}\langle1,0\vert\Psi\rangle_{ABC}
\propto
\ket{\psi(-g\alpha)}_B.
\eeq

In the case where the channel between Alice and Bob 
is subject to a linear loss with the rate $R_E$, 
one can consider an external mode E. 
Bob chooses the cat-state amplitude as 
\beq\label{beta in the losssy case}
\beta =\sqrt{ \frac{1-R_A}{R_A R_B (1-R_E)} } \alpha. 
\eeq
The whole state before Alice's measurement is 
\beqa\label{Psi_ABCE}
&&\ket{\Psi}_{ABCE}
\nonumber\\
&&=
\hat V_{AC} \ket{\psi}_A
\hat V_{EC} \hat V_{BC} \ket{\Phi}_B \ket{0}_C \ket{0}_E
\nonumber\\
&&=
{\cal{N}}c_0
\ket{ 0 }_A
\ket{ g\alpha }_B
\ket{-\frac{1}{\sqrt{R_A}}  \alpha }_C
\ket{-\varepsilon }_E
\nonumber\\
&&-
{\cal{N}}c_0
\ket{ 2\sqrt{1-R_A} \alpha }_A
\ket{-g\alpha }_B
\ket{ \frac{1-2R_A}{\sqrt{R_A} } \alpha }_C
\ket{ \varepsilon }_E
\nonumber\\
&&+
{\cal{N}}c_1
\ket{-2\sqrt{1-R_A} \alpha }_A
\ket{ g\alpha }_B
\ket{-\frac{1-2R_A}{\sqrt{R_A} } \alpha }_C
\ket{-\varepsilon }_E
\nonumber\\
&&-
{\cal{N}}c_1
\ket{ 0 }_A
\ket{-g\alpha }_B
\ket{ \frac{1}{\sqrt{R_A}}  \alpha }_C
\ket{ \varepsilon }_E
\eeqa
with the gain  
\beq\label{gain in the lossy case}
g=\sqrt{ \frac{ (1-R_A) (1-R_B) }{ R_A R_B (1-R_E) }  }
\eeq 
and 
$\varepsilon=\sqrt{(1-R_A)R_E/R_A(1-R_E)}\alpha$.

%
\section{Extension to multi-ary coherent states}
\label{appB}
%

The binary case can be extended to $M$-ary phase-shift-keyed 
coherent states $\ket{\alpha_m}$, 
where 
\beq\label{alpha_m}
\alpha_m=\alpha u^m,\quad 
u=e^{ 2\pi i/M}. 
\eeq
Here $\alpha_0=\alpha$ is taken to be real. 
The states are generated as 
\beq\label{state_alpha_m}
\ket{\alpha_m}=\hat V^m \ket{\alpha_0} 
\eeq
by modulating the phase of the coherent state 
$\ket{\alpha}$ with 
\beq\label{V_def}
\hat V=\exp\left( \frac{2\pi i}{M} \hat n \right). 
\eeq
An input state at Alice is generally a superposition state 
\beq\label{M-ary input psi}
\ket{\psi}_A
=
\sum_{m=0}^{M-1} c_m \ket{\alpha_m}_A. 
\eeq
Bob prepares a cat state for the entanglement resource, 
\beq
\ket{\Phi}_B
=
\sum_{k=0}^{M-1} b_m \ket{\beta_m}_B 
\eeq
where $\beta_m=\beta u^m$ with $\beta$ real. 
In order to analyze the scheme and its performmance, 
we introduce the orthonormal basis $\{\ket{\omega_m}\}$ 
in the space spanned by $\{ \ket{\beta_m} \}$ 
as follows 
\beq\label{omega_m}
\ket{\omega_m}
=
\frac{1}{\sqrt{M\lambda_m}}
\sum_{k=0}^{M-1} u^{-mk} \ket{\beta_k} 
\eeq
where 
\beq
\lambda_m
=
\sum_{k=0}^{M-1} u^{-km} \amp{\beta_0}{\beta_k}. 
\eeq
The orthonormality $\amp{\omega_{m'}}{\omega_m}=\delta_{m',m}$ 
can be verified by a relation  
\beq\label{V_diagonalized}
\sum_{k=0}^{M-1} u^{(m-n)k}=M \delta_{m,n+lM} 
\,\,\, (\forall\text{ integer }l) 
\eeq
and that the Gram matrix $[\amp{\beta_{k'}}{\beta_k}]$ 
is cyclic. 
The coherent states can be expanded as 
\beq
\ket{\beta_m}
=
\frac{1}{\sqrt M}
\sum_{k=0}^{M-1} \sqrt{\lambda_k} u^{mk} \ket{\omega_k}. 
\eeq
Then one can see that
\beq
\hat\rho
=\sum_{m=0}^{M-1} \proj{\beta_m}
=\sum_{m=0}^{M-1} \lambda_m \proj{\omega_m}. 
\eeq
Thus $\lambda_m$ are the eigenvalues 
of the density operator of the ensemble $\{ \ket{\beta_m} \}$. 
The mean photon number of the basis states is 
\beq
\bra{\omega_m} \hat n \ket{\omega_m}
=\frac{\lambda_{m-1}}{\lambda_m}|\beta|^2. 
\eeq
To maximize the success probability of Alice's measurement, 
one should use the $\ket{\omega_m}$ 
which has the maximum photon number 
for the entanglement resource. 
For relatively smaller $|\beta|$, 
it is the $\ket{\omega_{M-1}}$. 
In fact, the basis states can be represented by 
the number states as 
\beq\label{omega_m by |n>}
\ket{\omega_m}
=
\sqrt{\frac{M}{\lambda_m}}e^{-\beta^2/2}
\sum_{l=0}^{\infty} \frac{\beta^{m+lM}}{\sqrt{(m+lM)!}} \ket{m+lM}.  
\eeq
Thus $\ket{\omega_m}$ consists of a set of the photon number 
states $\{ \ket{m+lM}; l=0,1,... \}$.

Now let us see the case of $M=4$ $(u=i)$. 
The basis states are explicitly given by 
\beqa
\ket{\omega_0} &=& \frac{2e^{-\beta^2/2}}{\sqrt{\lambda_0}} 
\Bigl(\ket{0}+\frac{\beta^4}{\sqrt{4!}}\ket{4}+\cdots\Bigr) 
\nonumber\\
\ket{\omega_1} &=& \frac{2e^{-\beta^2/2}}{\sqrt{\lambda_1}} 
\Bigl(\beta\ket{1}+\frac{\beta^5}{\sqrt{5!}}\ket{5}+\cdots\Bigr) 
\nonumber\\
\ket{\omega_2} &=& \frac{2e^{-\beta^2/2}}{\sqrt{\lambda_2}} 
\Bigl(\frac{\beta^2}{\sqrt{2!}}\ket{2}
+\frac{\beta^6}{\sqrt{6!}}\ket{6}+\cdots\Bigr) 
\nonumber\\
\ket{\omega_3} &=& \frac{2e^{-\beta^2/2}}{\sqrt{\lambda_3}} 
\Bigl(\frac{\beta^3}{\sqrt{3!}}\ket{3}
+\frac{\beta^7}{\sqrt{7!}}\ket{7}+\cdots\Bigr) 
\eeqa
with the eigenvalues 
\beq
\begin{array}{lll}
\lambda_0 &=& 2 e^{-\beta^2} (\cosh\beta^2+\cos\beta^2), 
\\
\lambda_1 &=& 2 e^{-\beta^2} (\sinh\beta^2+\sin\beta^2), 
\\
\lambda_2 &=& 2 e^{-\beta^2} (\cosh\beta^2-\cos\beta^2), 
\\
\lambda_3 &=& 2 e^{-\beta^2} (\sinh\beta^2-\sin\beta^2). 
\end{array}
\eeq
The cat state for the entanglement resource is chosen as 
\beq
\ket{\Phi}_B
=\ket{\omega_3}_B
=
\frac{1}{\sqrt{4\lambda_3}}
\sum_{k=0}^{M-1} u^k \ket{\beta_k}_B. 
\eeq
The above state is beam-split into paths B and C, 
the component of mode C is sent to Alice through a lossy channel, 
and then 
combined with the input at path A.

Bob chooses the cat-state amplitude 
as Eq. (\ref{beta in the losssy case}). 
The whole state before the measurement is given by
\beqa
\ket{\Psi}_{BACE}
&=&
\sum_{m=0}^3
c_m
\sum_{k=0}^3
\frac{u^k}{\sqrt{4\lambda_3}}
\ket{g \alpha u^k}_B
\nonumber\\
&\otimes&
\ket{ \sqrt{1-R_A} \alpha (u^m-u^k) }_A
\nonumber\\
&\otimes&
\ket{ -\frac{\alpha}{\sqrt{R_A}}
      \Bigl[R_A u^m +(1-R_A)u^k \Bigr]  }_C
\nonumber\\
&\otimes&
\ket{\varepsilon u^k}_E
\eeqa
with the gain given by Eq. (\ref{gain in the lossy case}).

We set $R_A=0.5$. 
Alice further introduces additional modes A' and C' 
to implement the four-port single photon detection 
as shown in Fig. 1{\bf b}. 
The state before the detection is given by 
\beqa
\ket{\Psi}_{BAA'CC'E}
&=&
\sum_{m=0}^3
c_m
\sum_{k=0}^3
u^k \ket{g \alpha u^k}_B
\nonumber\\
&\otimes&
\ket{ \alpha \frac{u^m-u^k}{2} }_A
\nonumber\\
&\otimes&
\ket{-\alpha \frac{u^m+u^k}{2} }_C
\nonumber\\
&\otimes&
\ket{ \alpha \frac{u^m-u^{m+1}+u^k+u^{k+1}}{2\sqrt2} }_{A'}
\nonumber\\
&\otimes&
\ket{ \alpha \frac{u^m+u^{m+1}+u^k-u^{k+1}}{2\sqrt2} }_{C'}
\nonumber\\
&\otimes&
\ket{ \frac{R_E}{\sqrt{1-R_E}} \alpha u^k }_E
\eeqa

When the loss can be neglected ($R_E=0$), 
the input $\ket{\psi}$ of Eq. (\ref{M-ary input psi})
can be faithfully tele-amplified to the target state 
\beq\label{M-ary output psi}
\ket{g\psi}_B
=
\sum_{m=0}^{M-1} c_m \ket{g\alpha_m}_B, 
\eeq
by selecting a set of Alice's measurement result 
as (A, A', C, C')=(0,1,1,1), 
namely 
no count at port A while single-photon counts at 
port A', C, and C'.

In the lossy case, 
it is impossible to teleport a superposition state faithfully. 
However, when an input is restricted to a classical state 
drawn from the set $\ket{\alpha_m}$, 
then the tele-amplification to the target pure state is possible. 
Actually depending on a set of the results at the four ports,
(A, A', C, C'),
the inputs are tele-amplified as
\beq\label{QPSK tele-amplification0}
\begin{array}{llll}
\ket{\alpha_m} &\mapsto& \ket{g\alpha_m},
    \,\,&\mathrm{for}\,\, (0,1,1,1),
\\
\ket{\alpha_m} &\mapsto& \ket{ig\alpha_m},
    \,\,&\mathrm{for}\,\, (1,0,1,1),
\\
\ket{\alpha_m} &\mapsto& \ket{-g\alpha_m},
    \,\,&\mathrm{for}\,\, (1,1,0,1),
\\
\ket{\alpha_m} &\mapsto& \ket{-ig\alpha_m},
    \,\,&\mathrm{for}\,\, (1,1,1,0).
\end{array}
\eeq

%
\section{On/off detection at Alice}
\label{appC}
%

In our experiment, 
Alice's measurement is implemented by avalanche photodiodes (APDs) 
instead of ideal ``single-photon detectors" 
that discriminate between ``0", ``1" and ``2 or more" photons. 
APDs cannot, however, discriminate photon numbers, 
but distinguish merely the vacuum or non-vacuum state, 
i.e. ``off" or ``on". 
They are represented by the operators 
$\hat \Pi_\mathrm{off}=\proj{0}$ 
and 
$\hat \Pi_\mathrm{on}=\hat I - \proj{0}$. 
Then the tele-amplification described in the previous section 
should be corrected slightly. 
For example, 
Eq. (\ref{output in ideal tele-amplification:binary}) 
for the binary case becomes 
\begin{multline}
{}_{ABC}\langle\Psi\vert 
\hat \Pi_\mathrm{on}^A \hat \Pi_\mathrm{off}^C 
\vert\Psi\rangle_{ABC}
= \vert\psi(-g\alpha)\rangle_B\langle\psi(-g\alpha)\vert  \\
+ \tanh\Bigl(\frac{\tilde\alpha}{2}\Bigr)
\vert\tilde\psi(-g\alpha)\rangle_B\langle\tilde\psi(-g\alpha)\vert
\end{multline}
where 
$\vert\tilde\psi(-g\alpha)\rangle
=c_0 \ket{\alpha} - c_1 \ket{-\alpha}$ 
and 
$\tilde\alpha=2\sqrt{1-R_A}\alpha$. 
The second term is the correction. 
When $\alpha$ is small, 
the coefficient of the second term is small as 
$\tanh(\tilde\alpha/2)\sim\tilde\alpha/2$. 
In this regime, 
the tele-amplification would approximately work with on/off detection. 
However,
in general, 
the second term cannot be ignored.

If the input state was a coherent state, 
i.e., $c_0 = 0$ or $c_1 = 0$, 
the above state would become a pure coherent state with the gain $g$, 
and on/off detection would be sufficient.

%
%
%

%
\section{Success probability}
\label{appD}
%

%
\subsection{Loss tolerant quantum relay}
%

In the binary case, the input is either of 
$\ket{\alpha}$ and $\ket{-\alpha}$. 
The whole state before Alice's measurement is 
%
%
%
%
%
%
\beq\label{Psi_ABCE}
\begin{split}
\ket{\Psi_\pm}
&=
\hat V_{AC} \ket{\pm\alpha}_A
\hat V_{EC} \hat V_{BC} \ket{\Phi}_B \ket{0}_C \ket{0}_E
\\
&=
\pm
{\cal{N}}
\ket{ 0 }_A
\ket{\pm g\alpha }_B
\ket{\mp\frac{1}{\sqrt{R_A}}  \alpha }_C
\ket{\mp \varepsilon }_E\\
\mp
{\cal{N}}&
\ket{\pm 2\sqrt{1-R_A} \alpha }_A
\ket{\mp g\alpha }_B
\ket{\pm \frac{1-2R_A}{\sqrt{R_A} } \alpha }_C
\ket{\pm \varepsilon }_E 
\end{split}
\eeq
%
%
%
%
%
%
The success probability of the tele-amplification 
$\ket{\pm\alpha}_A \mapsto\ket{\pm g\alpha}_B$ 
is given by the expectation value of 
$
\hat \Pi_{10}\equiv
\hat \Pi_\mathrm{on}^A \otimes \hat \Pi_\mathrm{off}^C
$ 
as  
\beqa
P_\mathrm{Tele-amp}^{(2)}
&=&
\frac{1}{2}
\bra{\Psi_+} \hat \Pi_{10} \ket{\Psi_+}
+
\frac{1}{2}
\bra{\Psi_-} \hat \Pi_{10} \ket{\Psi_-}
\nonumber\\
&=&
\frac{
\exp\left[-\frac{(1-2R_A)^2}{R_A}\alpha^2\right]
-
\exp\left[-\frac{\alpha^2}{R_A}\right]
}
{
2\left(1-
\exp\left[-\frac{2(1-R_A)}{R_A R_B(1-R_E)}\alpha^2\right]
\right)
}.
\eeqa

In the case of 4-PSK states, 
the state before Alice's measurement is given by 
\beqa
\ket{\Psi_m}
&=&
\sum_{k=0}^3
u^k \ket{g \alpha u^k}_B
\nonumber\\
&\otimes&
\ket{ \alpha \frac{u^m-u^k}{2} }_A
\ket{-\alpha \frac{u^m+u^k}{2} }_C
\nonumber\\
&\otimes&
\ket{ \alpha \frac{u^m-u^{m+1}+u^k+u^{k+1}}{2\sqrt2} }_{A'}
\nonumber\\
&\otimes&
\ket{ \alpha \frac{u^m+u^{m+1}+u^k-u^{k+1}}{2\sqrt2} }_{C'}
\nonumber\\
&\otimes&
\ket{ \frac{R_E}{\sqrt{1-R_E}} \alpha u^k }_E
\eeqa
The success probability of  
$\ket{\alpha_m}_A \mapsto\ket{g\alpha_m}_B$ 
is given by the expectation value of 
\beq
\hat \Pi_{0111}
\equiv
\hat \Pi_\mathrm{off}^A \otimes 
\hat \Pi_\mathrm{on}^{A'}\otimes 
\hat \Pi_\mathrm{on}^C \otimes 
\hat \Pi_\mathrm{on}^{C'}
\eeq 
as
\beqa
P_\mathrm{Tele-amp}^{(4)}
&=&
\frac{1}{4}
\sum_{k=0}^{3}
\bra{\Psi_m} \hat \Pi_{0111} \ket{\Psi_m}
\nonumber\\
&=&
\bra{\Psi_0} \hat \Pi_{0111} \ket{\Psi_0}
\nonumber\\
&=&
\frac{
(1-e^{-\alpha^2/2})^2
(1-e^{-\alpha^2})
}
{
4
\lambda_3 (\frac{\alpha^2}{ R_B(1-R_E)})
}
\eeqa
where 
\beq
\lambda_3 (x) = 2 e^{-x} (\sinh x-\sin x). 
\eeq

\begin{figure}
\begin{center}
\includegraphics[width=0.85\linewidth]
{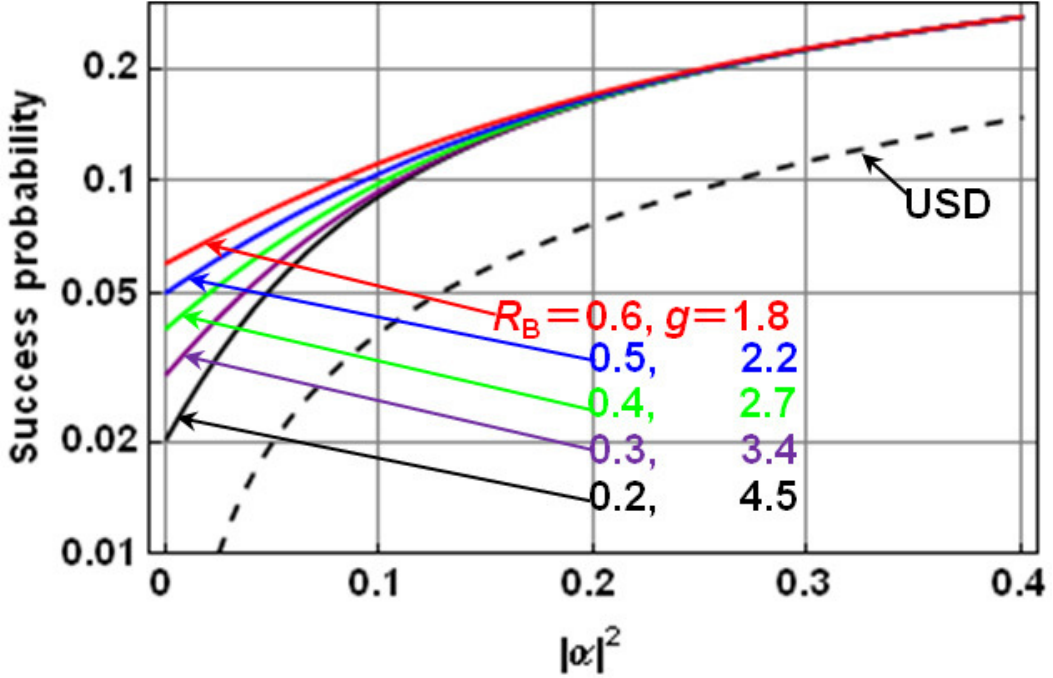} 
\caption{\label{Success_probability_BPSK}
Success probabilities for the case of the BPSK coherent states. 
The channel loss is assumed be 80\% ($R_E=0.8$). 
The solid lines are for the quantum relay with several 
cases of the amplification gains, 
while the dashed line is for the measure-resend 
strategy with the unambiguous state discrimination. 
}
\end{center}
\end{figure}

\begin{figure}
\begin{center}
\includegraphics[width=0.85\linewidth]
{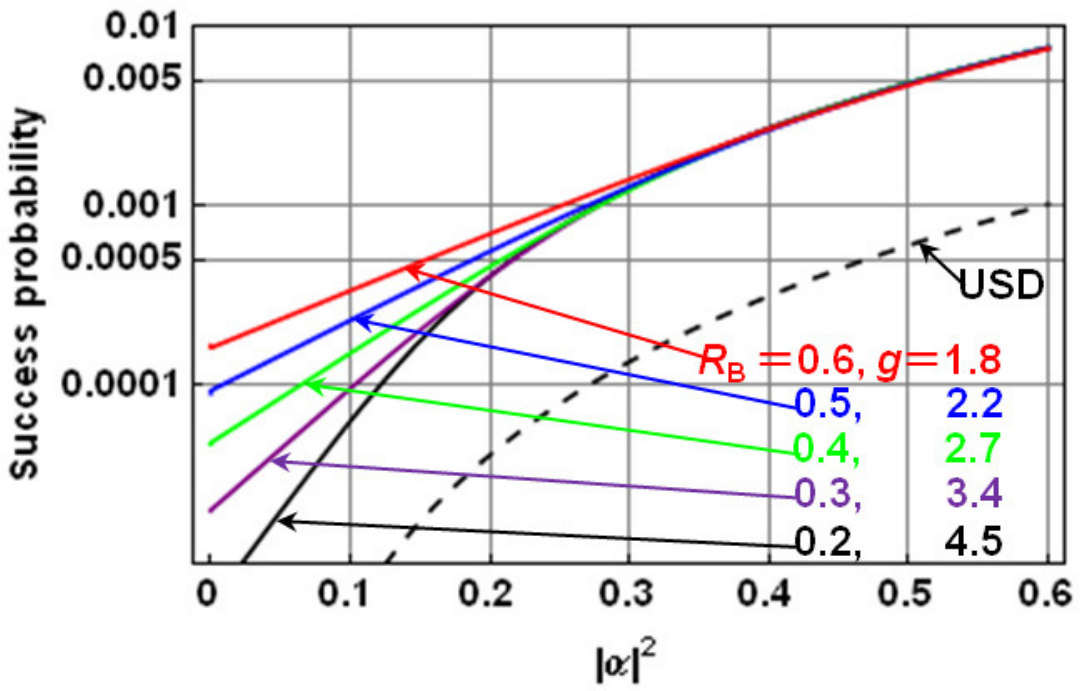}
\caption{\label{Success_probability_4PSK}
Success probabilities for the case of the 4PSK coherent states. 
The channel loss is assumed be 80\% ($R_E=0.8$). 
}
\end{center}
\end{figure}

%
\subsection{Measure-resend strategy}
%

The task to relay attenuated coherent states to the receiver, 
converting them faithfully to the target amplified states, 
can also be realized by a classical strategy. 
A typical one is a measure-resend strategy. 
In the intermediate node, 
Bob has attenuated states 
$\{ \ket{\sqrt{1-R_E} \alpha_m} \}$. 
He tries to discriminate them unambiguosly without errors, 
but at a finite success rate, 
referred to as unambiguous state discrimination (USD), 
and then 
prepare a target amplified state $\ket{g \sqrt{1-R_E} \alpha_m}$ 
for the measurement result $m$. 
The success rate is well known 
for this kind of equally probable symmetric states 
\cite{Chefles_Barnett_PLA98}. 
Denoting 
$\ket{\gamma_m} = \ket{\sqrt{1-R_E} \alpha_m}$ 
and 
using the eigenvalues and the diagonalizing vectors 
of the density operator 
\beq
\hat\rho
=\sum_{m=0}^{M-1} \proj{\gamma_m}
=\sum_{m=0}^{M-1} \lambda_m \proj{\omega_m}, 
\eeq
the success rate is given by 
\beq
P_\mathrm{USD}=\min_k{\lambda_k}. 
\eeq
The signal states are represented as 
\beqa
\ket{\gamma_m}
&=&
\ket{\sqrt{1-R_E} \alpha_m}
\nonumber
\\
&=&
\frac{1}{\sqrt{M}}
\sum_{k=0}^{M-1} \sqrt{\lambda_k} u^{mk} \ket{\omega_k}. 
\eeqa
The detection operators are given by 
\beq
\hat\Pi_m=\frac{\Lambda}{M} P_{USD}
\proj{\gamma_m^\bot}
\eeq
for the signal state $\ket{\gamma_m}$, 
using the reciprocal states 
\beq
\ket{\gamma_m^\bot}
=
\frac{1}{\sqrt{\Lambda}}
\sum_{k=0}^{M-1} \frac{u^{mk}}{\sqrt{\lambda_k}} \ket{\omega_k}  
\eeq
where $\Lambda=\sum_k \lambda_k^{-1}$. 
They satisfy the orthogonality relation 
\beq
\amp{\gamma_m^\bot}{\gamma_{m'}}
=
\sqrt{\frac{M}{\Lambda}}\delta_{m,m'}.  
\eeq
The operator for the inconclusive result is given by 
\beq
\hat\Pi_F=\hat I - 
\sum_{m=0}^{M-1} \hat\Pi_m.  
\eeq

\begin{figure*}
\begin{center}
\includegraphics[width=.98\linewidth]
{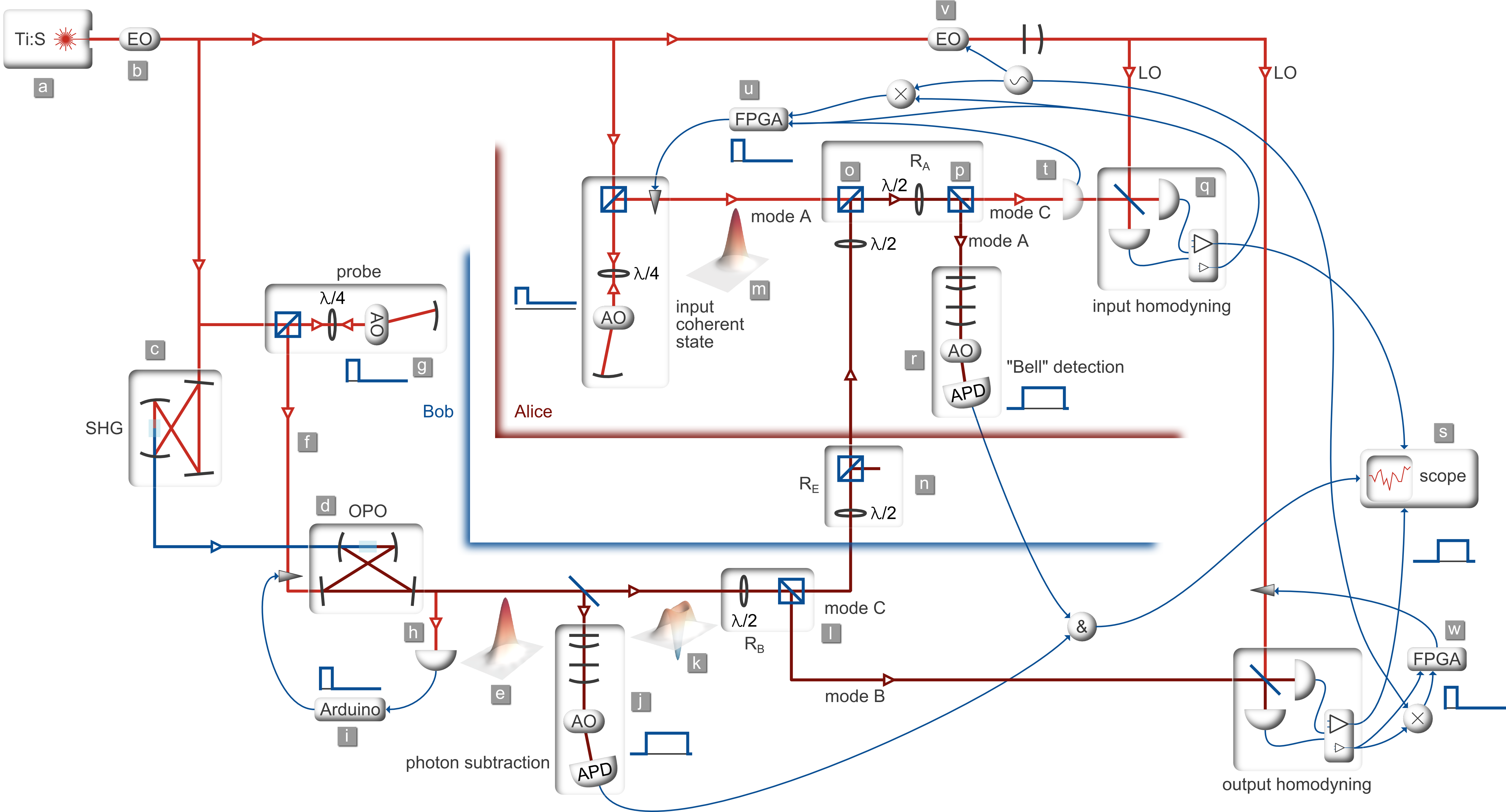}   %
\caption{\label{exp_setup}
Sketch of the experimental setup for tele-amplification of
a coherent state from Alice to Bob using shared entanglement
in the form of a photon-subtracted squeezed vacuum state.
}
\end{center}
\end{figure*}

%
\subsection{Numerical results}
%

Numerical results of the success probabilities 
in the case of $R_E=0.8$ (80\% loss) 
are shown 
in Fig. \ref{Success_probability_BPSK} 
for the BPSK coherent states, 
and 
in Fig. \ref{Success_probability_4PSK}
for the 4PSK coherent states, 
in comparison with that by the measure-resend strategy. 
The quantum relay can attain higher success rates 
than that by the above measure-resend strategy. 
In the case of 4PSK states, 
the difference is as high as ten times.

%
\section{Experimental details}
\label{appE}
%

\subsection{Setup}

The most relevant elements of the experimental
setup are sketched in Fig. \ref{exp_setup} 
and described in the following.

\subsubsection*{Resource state generation}

The output of an 860 nm continuous-wave Ti:Sapph laser (a) is
phase-modulated by an EOM (b) for Pound-Drever locking of the
SHG and OPO cavities. Parts of the beam are tapped off for use
as local oscillators, coherent input beam, probe beam and
cavity locking beams, but the main part is frequency-doubled
in the second harmonic generator (c). The 430 nm output
of this SHG pumps a bow-tie configuration optical parametric
oscillator (OPO, d) with a PPKTP crystal and a HWHM bandwidth
of $\gamma/2\pi = 4.5\ \mathrm{MHz}$. The down-converted
light leaving the cavity is in a squeezed vacuum state (e).
A probe beam (f) is injected into the OPO for the purpose of
locking phases and filtering cavities further downstream. 

The whole experiment is running in an alternating lock/measure
cycle at a rate of 10 kHz. The probe beam is thus switched on 
and off by double-passes through two acousto-optical modulators
(only one shown in the figure) that are driven for only 20\% of
the 10 kHz cycle time, as indicated by the little blue pulse
diagram (g) (this diagram is repeated for other relevant parts
of the setup).

In order to act as a phase-reference for the squeezing, the
probe beam is locked in phase with the squeezed quadrature
by observing its classical parametric
amplification in the OPO through a $<$1\% tap-off of the OPO
output (h). To obtain an error-signal, the probe's phase is
dithered on a piezo-mounted mirror by a micro-controller 
unit (Arduino) which also processes the detected signal and 
provides feedback to lock the phase (i).

A photon is subtracted from the squeezed vacuum at random
times by the detection on an avalanche photo-diode (APD)
detector, placed after a 5\% tapping beam-splitter and two
frequency-filtering cavities (j). The APD is protected from
the strong probe beam by an AOM that directs the OPO output
to the APD only during the intervals when the probe beam is
switched off. The resulting photon-subtracted
squeezed vacuum (PSSV) state (k) is the resource of entanglement
in our protocol, after it is split into two modes propagating
towards the Alice and Bob sections of the setup. In the
description of the protocol in the main section, a fraction
$R_B=0.1$ is reflected towards Alice. In the actual experiment
(and in the setup sketch), Alice's fraction $R_B$ is actually
taken as the transmitted part of a variable beam-splitter fixed
at 90\% reflection (l). Bob's share of the entangled state is
directed towards a homodyne detector for output state analysis.

\begin{figure}
\begin{center}
\includegraphics[width=0.8\linewidth]
{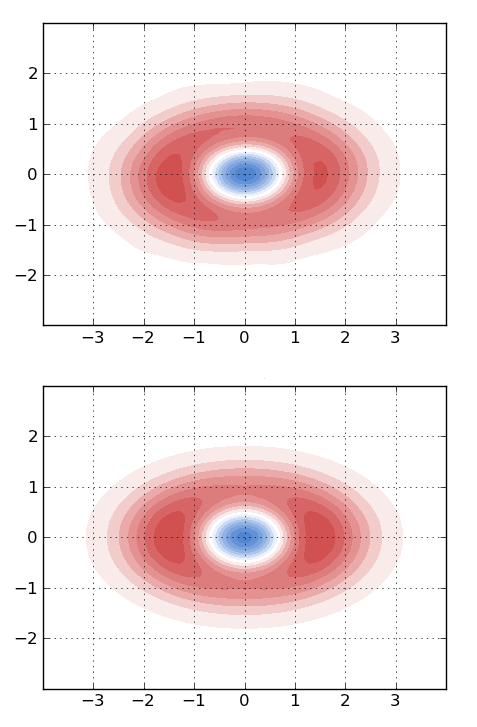}   %
\caption{\label{fig:resourcecat_wigner}
Wigner function for one of the PSSV states used as approximations
for the odd cat resource state, in this case with $\epsilon=0.20$. 
The upper is experimentally generated and tomographically 
reconstructed, while the lower one is obtained from our model.
The fidelity between them is above 98\%, showing the validity of
the model.
}
\end{center}
\end{figure}

\begin{figure}
\begin{center}
\includegraphics[width=0.95\linewidth]
{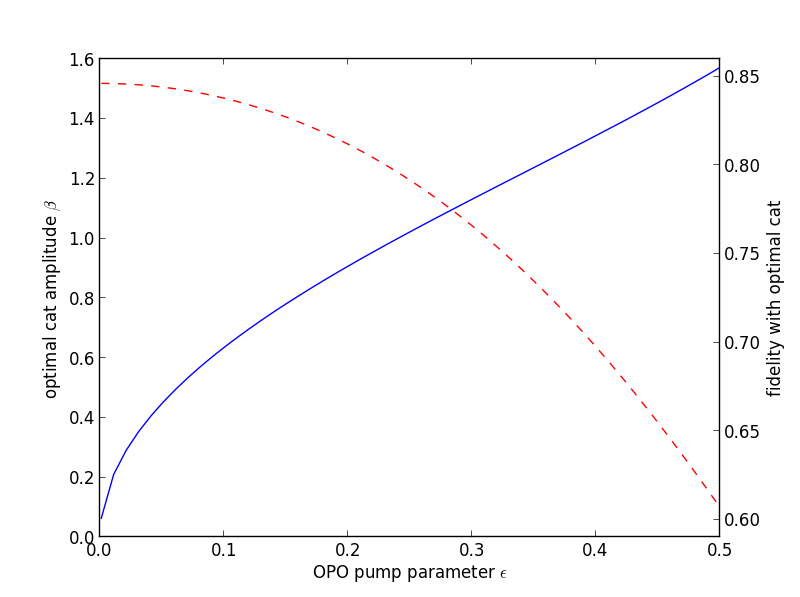}   %
\caption{\label{fig:eps2beta}
Relation between OPO pump parameter $\epsilon$
for the production of realistic PSSV
states and the amplitude of the cat state $\ket{\Phi_-(\beta)}$
that maximizes the mutual fidelity. The blue curve indicates
these optimal $\beta$ amplitudes, while the dashed red curve
shows the corresponding fidelities. In the experiment, we used
$\epsilon$ in the range 0.15--0.31.
}
\end{center}
\end{figure}

The amount of squeezing produced by the OPO determines the
amplitude of the cat-like PSSV state. It is regulated by the
pump parameter $\epsilon=\sqrt{P_\mathrm{pump}/
P_\mathrm{threshold}}$. To find what cat amplitude $\beta$
a given pump parameter corresponds to, we model the PSSV as
in Ref. [23] of the main text. Here, we include the OPO temporal
correlations, the temporal modes of the APD and homodyne detection
(described in the following section), the 96\% escape efficiency
of the OPO, the 95\% propagation efficiency towards the 
beam-splitter (l), the 5\% tapping ratio, the $\sim$10\% overall 
APD detection efficiency, and the filtering bandwidth to find a
$\hat{\rho}_\mathrm{PSSV}$ that closely emulates the actually
produced states. Fig. \ref{fig:resourcecat_wigner} shows an
example of an experimentally generated PSSV state and, for comparison,
the modelled state with equivalent parameters.
We then maximize for $\beta$ the state's fidelity with a true 
cat state, 
$\bra{\Phi_-(\beta)} \hat{\rho}_\mathrm{PSSV} \ket{\Phi_-(\beta)}$,
and get the $\epsilon \rightarrow \beta$ correspondence plotted
in Fig. \ref{fig:eps2beta} and thereby the
$\beta$ values of Table 1 in the main text.

\subsubsection*{Input state and teleportation}

Alice's input coherent state (m) is prepared in a configuration
of two double-pass AOMs similar to that used for the OPO probe beam,
with a strong phase locking beam switched on during the 20\%
locking part of the 10 kHz cycle.
However, instead of switching the light completely off during
the remaining 80\% of the cycle, a weak amplitude beam is 
generated instead. This is done by switching to a lower
voltage RF driving signal for the AOM. 

The share of the entangled PSSV state propagating from Bob to Alice
in mode C is optionally subjected to a loss at a variable
beam-splitter (n) before it is overlapped with her input state
on a polarizing beam splitter (PBS) in orthogonal polarizations (o). 
A half-waveplate followed by another PBS (p) then interferes the
two modes as the protocol's $R_A$ reflectivity beam-splitter.
When characterizing the input state, $R_A$ is set to 1, which
means that all of the input state is sent towards the homodyne
detector (q). Otherwise, when running the tele-amplification
protocol, $R_A$ is set to its appropriate value, and the
output of the beamsplitter in mode A is sent towards an APD (r) with the
same frequency filtering and chopping configuration as that
in (j).

The detection events from the two APDs are correlated with digital
timing electronics that pick out simultaneous events and triggers the
acquisition of Bob's homodyne signal at a fast digital oscilloscope (s).
The detected photo-currents are subsequently temporally filtered on a PC
to extract the measured quadrature values, as described in the following
section.

\subsubsection*{Phase locking}

For the tele-amplification protocol to work, the input states
$\ket{\pm\alpha}$ should be interfered in-phase with the anti-squeezed
quadrature of the entangled PSSV state. 
We do this by putting a normal intensity detector (t) at the mode C output
port of the beamsplitter instead of Alice's
homodyne detector (q). The detected interference signal between the 
probe portions of the input coherent state beam and the PSSV state beam
is used as the input to an 
FPGA-based lock unit, which provides feedback to the phase of the input 
beam (u). When the two beams are interfered at 90$^{\circ}$, we get the
desired phase relation, since the PSSV probe was locked to the squeezed
quadrature.

The phase of the local oscillators (LO) in the two homodyne detectors
can be locked to arbitrary phases relative to the PSSV probe by using a combination of DC and
side-band detection of the interference between the LO and the probe
beam. An 8 MHz phase modulation is applied to the LO (v), and the
interference signals observed by the fast homodyne detectors 
(from a separate low-gain amplification output) are demodulated at
that same frequency. This provides an interference signal that is 
90$^{\circ}$ out of phase with the DC signal. In the FPGA lock
units (u,w), the two signals are added with weighting factors
corresponding to the desired LO phase in phase space, resulting in
an error signal for the feedback to piezo-mounted mirrors in the LO
beam path (in the case of Bob's output homodyner) or in the input
beam path (for Alice).

All phase locks are engaged only during the intervals of the 10 kHz
experiment cycle in which the probe beams are turned on. For the
remaining time, the feedback signals are just held at their last actively
set value.

\subsection{Quantum state tomography}

\subsubsection*{Temporal modes}

The squeezed vacuum has a bandwidth given by the OPO's HWHM of
$\gamma/2\pi = 4.5\ \mathrm{MHz}$. Conditioned on an APD click at time $t_0$, 
the continuous-wave squeezed vacuum is converted into a 
temporally localized PSSV state in a temporal
mode around $t_0$ which, in the low-squeezing limit, has the form\cite{Molmer2006}
$\exp(-\gamma|t-t_0|)$. The filter cavities in front
of the APD, needed to remove the photons down-converted into the
many non-degenerate OPO resonances, modify the temporal mode to be
\begin{equation} \label{filtered_mode_function}
f(t) \propto \gamma^{-1}e^{-\gamma|t-t_0|} - \kappa^{-1}e^{-\kappa|t-t_0|} ,
\end{equation}
where $\kappa/2\pi \approx 25\ \mathrm{MHz}$ is the combined bandwidth of the
two filters, approximated by a single Lorentzian spectral profile.
This will also be the temporal mode of the teleported output state in the 
low-squeezing limit. 
For the state tomography, we therefore extract a single quadrature value 
from the continuous photo current signal of the homodyne detector 
by integrating it over a mode $f_{\mathrm{HD,out}}(t)$ equal to the one in
Eq. (\ref{filtered_mode_function}).
At higher squeezing levels, the optimal mode function is not that simple
\cite{Nielsen2007}, but in this work we stick to the simple expression
for all squeezing levels.

An interesting, but also complicating aspect of our current implementation
of the tele-amplification protocol is that the input and output states
are in rather different spectral modes: The input coherent state is derived
directly from the narrow-band laser, whereas the entangled PSSV state is in the
broadband mode described above. At a first glance it would appear like the two
modes will not interfere well and the teleportation will fail. However,
the spectral response of Alice's APD is very broad, so it is unable to
distinguish the modes. Thereby it can be said that the detection itself
induces the interference between the input and the entangled state. Another way
to see it is in the time domain: compared to the cw input beam and the
$\sim 1/\gamma$ extent of the PSSV state,
the temporal response of the APD ($\sim$350 ps jitter) is essentially
delta function-like. Within this short time window, almost no phase shift
will occur between the different frequency components, so interference
will not be destroyed.

One problem we do get from this spectral mismatch, however, is the issue
of which temporal mode, $f_{\mathrm{HD,in}}(t)$ to use for the definition of the coherent input state.
As the beam is continuous, the choice of temporal mode can be done 
arbitrarily. The photon number $n_{\mathrm{in}}$ within the chosen mode will be proportional
to the width of the mode function, so to obtain a desired coherent state
the intensity of the beam should be adjusted inversely proportional to
that width. In our experiment we make the rather natural choice to use the
$f_{\mathrm{HD,out}}(t)$ mode, such that we observe the
same temporal mode in both the input and the output homodyne tomography.
The measured $\alpha$ and $\alpha'$ values in Fig. 2 of the main paper
are therefore directly comparable.

However, this mode is not the one detected by the APD. Since the APD
with its delta function-like response is preceded by filtering cavities
which act as delays for incoming fields, its temporal
mode can be approximately described as a single-sided exponential decay,
with time constant given by the filter bandwidth,
\begin{equation} \label{filtered_apd_mode_function}
f_{\mathrm{APD}}(t) \propto e^{-\kappa|t-t_0|} H(t_0-t) ,
\end{equation}
with $H(t)$ being the Heaviside step function. Because the PSSV
entangled state is temporally localized, as opposed to the input state, 
the ratio of the photon numbers of the two states,
$n_{\mathrm{in}} / n_{\mathrm{PSSV}}$ will be different within the
different modes $f_{\mathrm{HD,in}}(t)$ and $f_{\mathrm{APD}}(t)$. 
There is therefore a mismatch between the input state amplitude, $\alpha$,
that we expect to have and the amplitude actually seen by Alice's APD,
which is the one to induce the teleportation. Thus, the $\beta$ and $R_A$
values that we experimentally adjusted to match a given $\alpha$ were
actually not optimal, and this resulted in output-to-target fidelities that
were lower than we could have otherwise obtained. In a possible follow-up
experiment, it would be advisable to consider this issue of the input
state amplitude in more detail. Simulations indicate that with optimized
settings, fidelities could have reached 0.94--0.99.

\subsubsection*{State reconstruction}	

\begin{figure}
\begin{center}
\includegraphics[width=\columnwidth]
{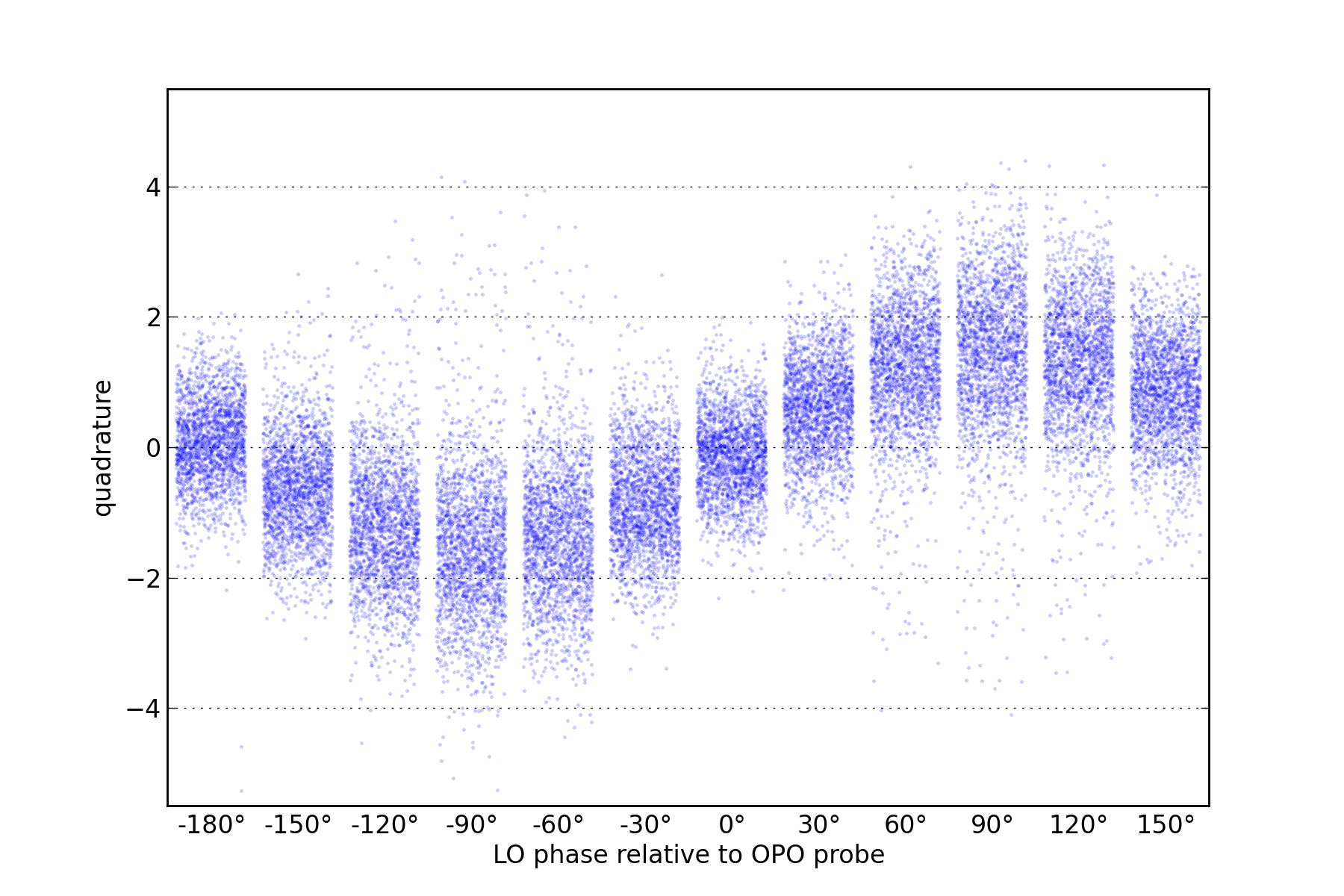}   %
\caption{\label{tomogram}
Example tomogram, showing the 12 $\times$ 2000 quadrature values of the
homodyne measurement of the output state in tele-amplification \#5.
}
\end{center}
\end{figure}

For a given realization of the tele-amplification, we construct a homodyne
tomogram of the output (and input) state by repeating the state preparation,
on/off detection and conditional homodyne detection multiple times, with the
LO phase of the homodyne detector fixed at various angles. After filtering
the oscilloscope traces with the chosen temporal mode, as described above,
the obtained quadrature values are normalized by vacuum traces recorded
under the same experimental conditions while we also pay attention to proper
offset correction of the traces, which can be particularly tricky for the
measurement of the input coherent state. That gives us a homodyne tomogram
like the one shown in Fig. \ref{tomogram}. From this we reconstruct an
estimate of the underlying quantum state using the maximum likelihood
method \cite{Lvovsky2004}. As mentioned in the Methods section, we correct
for the non-perfect detector efficiencies in order to get the most accurate
characterization of the protocol.

The phase values in the figure
indicate the relative phase between the local oscillator and the OPO-injected
probe beam. The probe beam is locked to the squeezed quadrature of the PSSV
state, and Alice's input coherent beam is locked at 90$^{\circ}$ to the probe
beam. Since we define our phase space in such a way that the anti-squeezing
is aligned along the $x$-axis and the input states have real amplitudes (i.e.
also along the $x$-axis), the 90$^{\circ}$ phase of the LO should correspond
to the $x$-quadrature. We therefore rotate the reconstructed quantum state
by $-90^{\circ}$ in phase space - the free choice of global phase.

%
\section{Modelling of qubit teleportation}
\label{appF}
%

To simulate the performance of our tele-amplifier setup in the case where
the input is an arbitrary coherent state qubit
\begin{equation}
\ket{\psi(\alpha,\theta,\phi)}  =
 \mathrm{cos}\frac{\theta}{2} \ket{\Phi_+(\alpha)}
+e^{ i \phi} 
   \mathrm{sin}\frac{\theta}{2} \ket{\Phi_-(\alpha)} ,
\end{equation}
we set up a model for the protocol, using Wigner function formalism.

As the input state to be teleported, we took a pure qubit state
of the above form. The initial resource state was a squeezed vacuum state
with appropriate squeezing levels. In the experiment, the squeezed
vacuum state within the homodyne-observed mode $f_\mathrm{HD,out}(t)$
is not pure. The impurity due to this mode selection can be modelled
quite well by propagating the initially pure squeezed vacuum through
a 92\% transmission beam-splitter. The losses suffered by the 
photon-subtracted squeezed vacuum were similarly modelled by virtual
beam-splitters, taking account of the 96\% escape efficiency of the OPO,
the 5\% tapping ratio for the photon subtraction, and the 95\% propagation
efficiency towards the separating beam-splitter. 
Alice's detector was modelled as an on/off detector with 10\% efficiency,
roughly corresponding to our APD's detection efficiency and the
transmission of the spectral filters.

\begin{figure}
\begin{center}
\includegraphics[width=\columnwidth]
{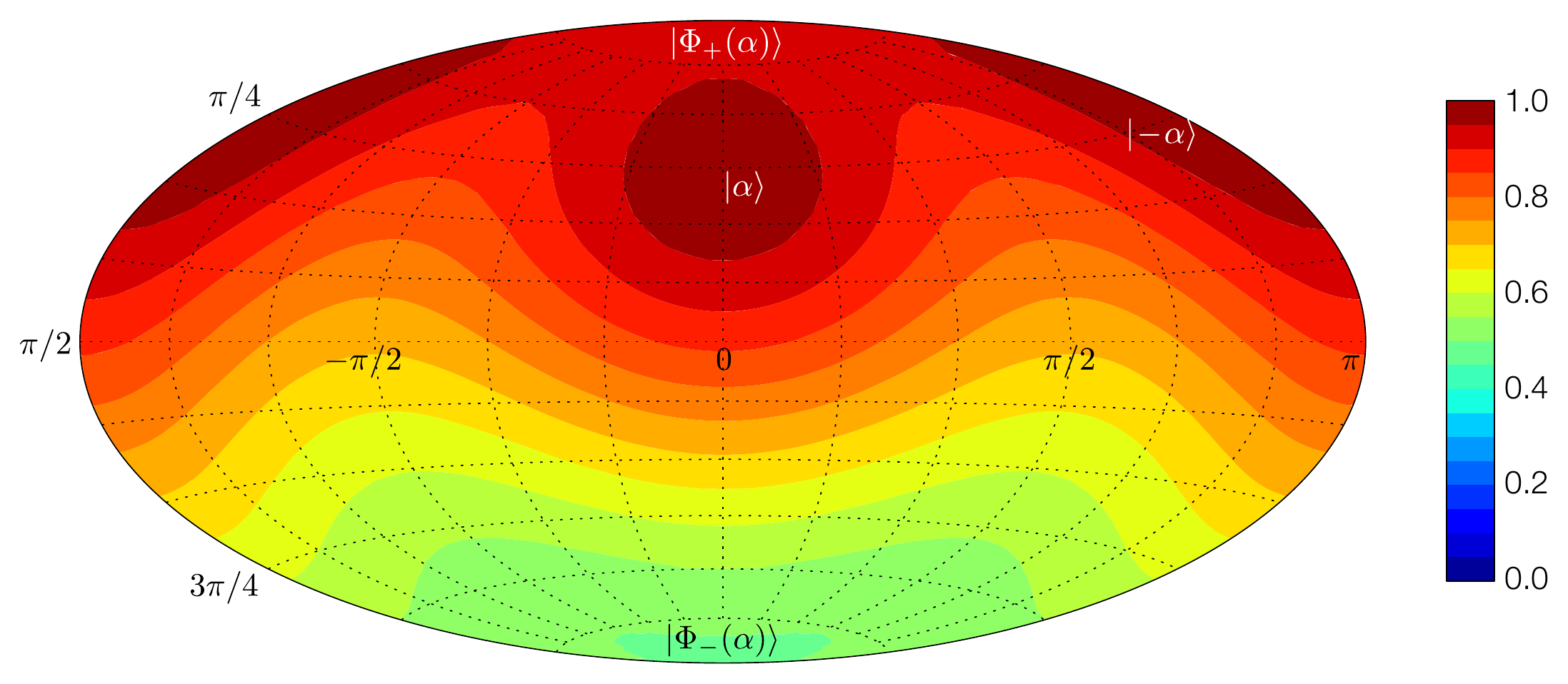}
\caption{\label{fig:fidelitymap}
Bloch sphere map of the fidelities between modelled and targeted outcomes of the
$\ket{\psi(\alpha,\theta,\phi)} \rightarrow \ket{\psi(\alpha',\theta,\phi)}$
tele-amplification, in this case for $\alpha=0.4$ and $\alpha'=0.6$. 
The averaged fidelity here is 77\%.
}
\end{center}
\end{figure}

For a given input amplitude $\alpha$ and desired output amplitude $\alpha'$,
we simulate the tele-amplification process for 168 evenly distributed
qubit states on the $(\theta,\phi)$ Bloch sphere and calculate the fidelity
between the output states and the targeted states 
$\ket{\psi(\alpha',\theta,\phi)}$. This results in a ``fidelity map'' like
the one in Fig. \ref{fig:fidelitymap} for every
$(\alpha,\alpha')$ setting. It is clear that the teleportation works best
for coherent state inputs (near 100\% fidelity) and for states near the
North Pole (which is the even cat) and not very well for states near the
South Pole (odd cat). By averaging over the Bloch sphere, we obtain an
average fidelity for the outcomes of the protocol for the given 
$(\alpha, \alpha')$ pair, giving one value for the average fidelity plot
in Fig. 3 of the main text.

%
\section{Application to quantum key distribution}
\label{appG}
%

The tele-amplification scheme would be useful to improve 
the performance of quantum key distribution (QKD) schemes 
which use phase-shifted coherent-state signals, 
such as B92 protocol with BPSK states 
\cite{Koashi2004,Tamaki2009} 
and BB84 protocol with 4PSK states 
\cite{Lo_Preskill2007_QIC_CohSt_BB84}. 
The tele-amplifications of BPSK and 4PSK states 
could be applied to B92 and BB84, respectively. 
The B92 with BPSK states would be more interesting 
from the viewpoint of practical implementation 
because the necessary cat-state resources are readily available 
in laboratories. 
Unfortunately, however, 
its security proof and performace evaluation 
when the tele-amplification is included are more involved. 
In contrast, 
BB84 protocol with 4PSK states can be analyzed more clearly
with the tele-amplification.

Therefore we consider the BB84 scheme 
where the key information is encoded 
in the relative phase of a coherent-state reference pulse 
and a coherent-state signal pulse as
\begin{subequations}\label{BB84 signals}
\begin{eqnarray}
\ket{\tilde{0}_X}&=&\ket{\alpha}_R\otimes\ket{\alpha}_A
\\
\ket{\tilde{1}_X}&=&\ket{\alpha}_R\otimes\ket{-\alpha}_A
\\
\ket{\tilde{0}_Y}&=&\ket{\alpha}_R\otimes\ket{i\alpha}_A
\\
\ket{\tilde{1}_Y}&=&\ket{\alpha}_R\otimes\ket{-i\alpha}_A
\end{eqnarray}
\end{subequations}
where mode A is for the signal pulse 
while mode R for the reference pulse 
\cite{Lo_Preskill2007_QIC_CohSt_BB84}. 
This scheme is referred to as the 4PSK-BB84 with reference pulse.  
The amplitude $\alpha$ is understood as the one 
at the receiver Bob. 
It reduces from $\alpha_\mathrm{in}$ 
at Alice by the channel loss as 
\beq
\alpha=\eta(L)\alpha_\mathrm{in}
\eeq
where
\beq\label{fiber transmission rate}
\eta(L)=10^{-\xi L/10}
\eeq
with the distance $L$ and the channel loss rate $\xi$. 
The phase of $\alpha$ is defined relative to 
a fixed classical phase reference frame that Eve can access.
Alice emits one of the four states. 
Bob randomly chooses one of two measurement apparatuses, 
the X-basis or the Y-basis measurement, and measures the signal. 
In the X-basis measurement, 
the two modes are first combined on a balanced beam splitter as
\begin{subequations}\label{BS for USD}
\begin{eqnarray}
\hat{V}\ket{\alpha}_R\ket{\alpha}_A
&=&
\ket{\sqrt{2}\alpha}_R\ket{0}_A
\\
\hat{V}\ket{\alpha}_R\ket{-\alpha}_A
&=&
\ket{0}_R\ket{-\sqrt{2}\alpha}_A
\end{eqnarray}
\end{subequations}
then directed to two on/off detectors described by operators
\begin{subequations}\label{On-Off POVM}
\begin{eqnarray}
\hat{\Pi}_\mathrm{off}
&=&
e^{-\nu}\sum^{\infty}_{m=0}(1-\eta_B)^m\ket{m}\bra{m}
\\
\hat{\Pi}_\mathrm{on}
&=&
\hat{I}-\hat{\Pi}_\mathrm{off}
\end{eqnarray}
\end{subequations}
where $\nu$ is the dark count probability 
and $\eta_B$ is the detection efficiency.
We define a POVM for making raw keys ``0" and ``1", 
and an inconclusive outcome ``2" by
\begin{subequations}\label{X-POVM}
\begin{eqnarray}
\hat{\Pi}_{0}^X
&=&
\hat{V}^\dagger\left(
\hat{\Pi}_\mathrm{on}^R\otimes\hat{\Pi}_\mathrm{off}^A 
+ \frac{1}{2}
\hat{\Pi}_\mathrm{on}^R\otimes\hat{\Pi}_\mathrm{on}^A
\right)\hat{V}
\\
\hat{\Pi}_{1}^X
&=&
\hat{V}^\dagger\left(
\hat{\Pi}_\mathrm{off}^R\otimes\hat{\Pi}_\mathrm{on}^A 
+ \frac{1}{2}
\hat{\Pi}_\mathrm{on}^R\otimes\hat{\Pi}_\mathrm{on}^A
\right)\hat{V}
\\
\hat{\Pi}_{2}^X
&=&
\hat{V}^\dagger
\hat{\Pi}_\mathrm{off}^R\otimes\hat{\Pi}_\mathrm{off}^A
\hat{V}.
\end{eqnarray}
\end{subequations}
We have three kinds of probabilities, 
$P_\mathrm{c}$ for correctly outputting 
the bit 0 (1) given the signal $\tilde{0}_X$ ($\tilde{1}_X$), 
$P_\mathrm{e}$ for incorrectly outputting 
the bit 0 (1) given the signal $\tilde{1}_X$ ($\tilde{0}_X$), 
and 
$P_\mathrm{i}$ for inconclusive outcome 
\beqa
P_\mathrm{c}
&=&
\bra {\tilde{0}_X} \hat{\Pi}^X_0 \ket {\tilde{0}_X}=
\bra {\tilde{1}_X} \hat{\Pi}^X_1 \ket {\tilde{1}_X}
\nonumber\\
&=&
\frac{1}{2}
\left(1-e^{-\nu-2\eta_B|\alpha|^2}\right)
(1+e^{-\nu})
\\
P_\mathrm{e}
&=&
\bra {\tilde{0}_X} \hat{\Pi}^X_1 \ket {\tilde{0}_X}=
\bra {\tilde{1}_X} \hat{\Pi}^X_0 \ket {\tilde{1}_X}
\nonumber\\
&=&
\frac{1}{2}
\left(1+e^{-\nu-2\eta_B|\alpha|^2}\right)
(1-e^{-\nu})
\\
P_\mathrm{i}
&=&
\bra {\tilde{0}_X} \hat{\Pi}^X_2 \ket {\tilde{0}_X}=
\bra {\tilde{1}_X} \hat{\Pi}^X_2 \ket {\tilde{1}_X}
\nonumber\\
&=&
e^{-2\nu-2\eta_B|\alpha|^2}. 
\eeqa
Similarly, the Y-basis measurement is described by a POVM
\begin{subequations}\label{Y-POVM}
\begin{eqnarray}
\hat{\Pi}^Y_0
&=&
e^{i\frac{\pi}{2}\hat{n}_A}
\hat{V}^{\dagger}
\Bigl(
\hat{\Pi}^R_\mathrm{on} \otimes \hat{\Pi}^A_\mathrm{off}
\nonumber\\
&&+
\frac{1}{2}\hat{\Pi}^R_\mathrm{on} \otimes \hat{\Pi}^A_\mathrm{on}
\Bigr)
\hat{V}e^{-i\frac{\pi}{2}{\hat{n}_A}}
\\
\hat{\Pi}^Y_1
&=&
e^{i\frac{\pi}{2}\hat{n}_A}
\hat{V}^{\dagger}
\Bigl(
\hat{\Pi}^R_\mathrm{off} \otimes \hat{\Pi}^A_\mathrm{on}
\nonumber\\
&&+
\frac{1}{2}\hat{\Pi}^R_\mathrm{on} \otimes \hat{\Pi}^A_\mathrm{on}
\Bigr)
\hat{V}e^{-i\frac{\pi}{2}{\hat{n}_A}}
\\
\hat{\Pi}^Y_2
&=&
e^{i\frac{\pi}{2}\hat{n}_A}
\hat{V}^{\dagger}
\hat{\Pi}^R_\mathrm{off} \otimes \hat{\Pi}^S_\mathrm{off}
\hat{V}e^{-i\frac{\pi}{2}{\hat{n}_A}}. 
\end{eqnarray}
\end{subequations}
where the factor $e^{i\frac{\pi}{2}{\hat{n}_A}}$ 
is for shifting the phase.
Eliminating the inconclusive outcomes, 
the filtered fraction for sifted keys is defined by 
\beq\label{Filtered fraction}
Q=1-P_\mathrm{i}
\eeq
and the bit error rate (BER) is defined by 
\beq\label{BER}
\delta=\frac{P_\mathrm{e}}{Q}
\eeq
Then the upper bound for the phase error rate (PER) is given by
\beqa\label{PER}
\delta_\mathrm{ph}
&=&
\delta+4{\Delta'}(1-{\Delta'})
(1-2\delta)
\nonumber\\
&+&
4(1-2{\Delta'})
\sqrt{{\Delta'}(1-{\Delta'})
\delta(1-\delta)}
\eeqa
where 
\begin{subequations}\label{imbalance of coin}
\begin{eqnarray}
\Delta'&=&\frac{\Delta}{Q}
\\
\Delta&=&\frac{1}{2}
\left[
1-e^{-\alpha_\mathrm{in}^2}
\left( \cos\alpha^2_\mathrm{in}+\sin\alpha^2_\mathrm{in} \right)
\right]. 
\end{eqnarray}
\end{subequations}
The quantity $\Delta$ specifies the imbalance of 
the ``coin" for the choice of X and Y bases depending 
on the states overlap among
$\{ 
\ket{\alpha_\mathrm{in}}, 
\ket{-\alpha_\mathrm{in}}, 
\ket{i\alpha_\mathrm{in}}, 
\ket{-i\alpha_\mathrm{in}} 
\}$.
The secure key generation probability is given by 
\beq\label{key rate of BB84}
G =\frac{1}{2}
Q\left[1-H(\delta)-H(\delta_\mathrm{ph})\right]
\eeq
where $H(x)$ is the binary Shannon entropy
\beq
H(x)=-x\log_2 x-(1-x)\log_2 (1-x).
\eeq

Now let us consider an extention of this BB84 implementation 
to a tele-amplification assisted scheme. 
The signals sent by Alice are 
$\alpha_m^\mathrm{in}=\alpha_\mathrm{in}u^m$ 
where $u=i$ and $m=$0, 1, 2 and 3. 
The signals first arrive at the relay node, 
referred to as Amy, 
and are then relayed to the receiver at the terminal node, 
referred to as Bob. 
The total distance between Alice and Bob is $L$. 
The relay node Amy is located at the distance $xL$ from Alice,
where $0<x<1$. 
The input coherent-state amplitude to the tele-amplifier at Amy 
is $\alpha_m= \sqrt{\eta(xL)}\alpha_\mathrm{in}u^m$. 
Bob prepares the resource cat state, beam-splits it, 
and sends Amy one half of the split cat-state 
over the distance $(1-x)L$. 
At the relay node, Amy combines it 
with the signal state $\ket{\alpha_m}_A$ on the beam splitter, 
and measures them by the four-port interferometric receiver. 
For simplicity, 
we assume that the single photons can be detected with 
perfect efficiency at the four ports. 
Three-photon coincidence counts at these ports 
herald the successful tele-amplification events. 
The dark count effect can be negligible 
by this multi-photon coincidence filtering. 
Bob finally receives the tele-apmplified signal 
$\ket{g\alpha_m}_B$, 
which is subject to the X or Y basis measurement. 
The gain is now a function of $x$ and $L$ 
\beq\label{gain(x,L)}
g(x,L)=\sqrt{\frac{1-R_B}{R_B \eta[(1-x)L]}}.
\eeq

Then the security proof in ref. 
\citenum{Lo_Preskill2007_QIC_CohSt_BB84} 
can be applied to the tele-amplification assisted BB84 
provided that the reference pulse arrives at Bob 
so as to be in
$\ket{g(x,L)\sqrt{\eta(xL)}\alpha_\mathrm{in}}_R
 \otimes
 \ket{g(x,L)\sqrt{\eta(xL)}\alpha_\mathrm{in}u^m}_B$.
The filtering fraction $Q$, the BER $\delta$, 
the PER $\delta_\mathrm{ph}$ and $\Delta'$ 
are given by replacing ${\alpha}$ in 
Eqs. (\ref{Filtered fraction}), (\ref{BER}), (\ref{PER}), 
and (\ref{imbalance of coin}) 
with the new one 
$g(x,L)\sqrt{\eta(xL)}\alpha_\mathrm{in}$.
Here note that the signal attenuation 
occurs only for the channel interval between Alice and Amy 
over a distance $xL$.
In the remaining channel with a distance $(1-x)L$, 
the signal attenuation is compensated 
by the cat-assisted amplification with the gain $g(x,L)$.

\begin{figure}[H]	
\begin{center}
\includegraphics[width=0.97\linewidth]
{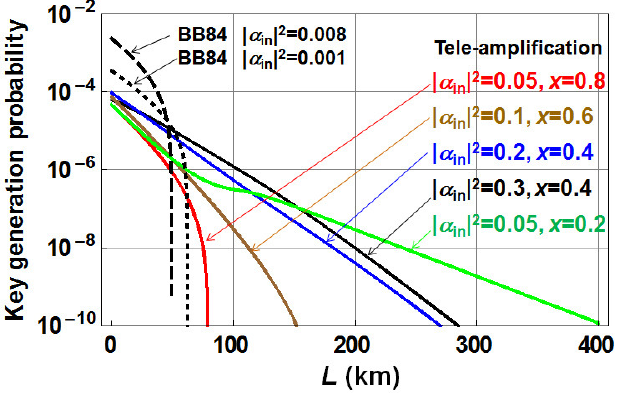}
\caption{\label{fig:Key_gen_prob}
The secure key generation probabilities as a function of distance. 
The dashed and dotted lines correspond to 
BB84 without tele-amplification. 
The solid lines correspond to tele-amplification assisted BB84 
where $R_B=0.2$. 
For all the cases, 
the dark count probability is $\nu=10^{-6}$, 
and the detection efficiency of the receiver Bob is $\eta_B=0.2$. 
}
\end{center}
\end{figure}

The secure key generation probability is finally given 
by multiplying the expression in Eq. (\ref{key rate of BB84}) 
by the success probability of the tele-amplification as
\beq
G= \frac{1}{2}P_\mathrm{Suc}Q
\left[1-H(\delta)-H(\delta_\mathrm{ph})\right]. 
\eeq
Some numerical results are shown in 
Fig. \ref{fig:Key_gen_prob} 
as a function of the transmission distance. 
In the following, 
the dark count probability is $\nu=10^{-6}$, 
and the detection efficiency of the receiver Bob is $\eta_B=0.2$. 
The dashed and dotted lines correspond to 
the 4PSK-BB84 with reference pulse. 
In this PSK coherent-state scheme, 
an eavesdropper Eve can effectively perform 
the photon-number splitting attack with phase information. 
So the input coherent-state amplitude should be set small. 
In Fig. \ref{fig:Key_gen_prob},  
the dashed and dotted lines correspond to 
$|\alpha_\mathrm{in}|^2=0.008$ and 0.001, respectively. 
The key generation probabilities decrease more rapidly than 
that of the phase-randomized decoyed scheme. 
The solid lines represent the performances 
of the tele-amplification assisted BB84 with $R_B=0.2$. 
It can be seen that 
the secure key generation probabilities  
are smaller than those without tele-amplification 
at short distances, 
however, they can remain at reasonable levels up to 
longer distances. 
The input coherent-state amplitude $\alpha_\mathrm{in}$ 
is allowed to be larger in the tele-amplification assisted BB84. 
The red and brown lines are the cases where 
the relay node Amy is located closer to Bob, 
namely $x=0.8$ and $x=0.6$, respectively. 
The blue and black lines are the cases where 
Amy is located at $0.4L$ from Alice, 
with 
$|\alpha_\mathrm{in}|^2=0.2$ and 0.3, respectively. 
The green line is the case where 
Amy is located much closer to Alice, namely $x=0.2$ 
with 
$|\alpha_\mathrm{in}|^2=0.05$.

\begin{figure}
\begin{center}
\includegraphics[width=0.92\linewidth]
{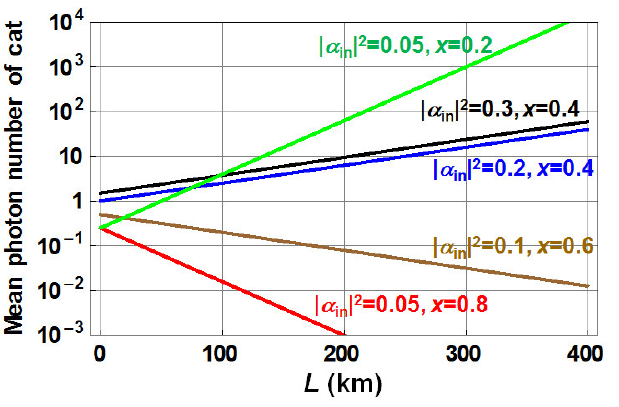}
\caption{\label{fig:Photon_number_cat}
The mean photon numbers of the required cat states 
as a function of distance. 
}
\end{center}
\end{figure}

The performance of the green line is remarkable, 
however, one should prepare the resource cat-state 
with a much larger amplitude. 
The mean photon number of the resource cat state is given by 
\beq
\beta(x,L)^2
=\frac{\eta(xL)\alpha_\mathrm{in}^2}{\eta[(1-x)L] R_B}. 
\eeq
It is shown in Fig. \ref{fig:Photon_number_cat}. 
To extend the distance beyond 200\,km, 
the resource cat-state should include more than a hundred photons.

\begin{figure}
\begin{center}
\includegraphics[width=0.92\linewidth]
{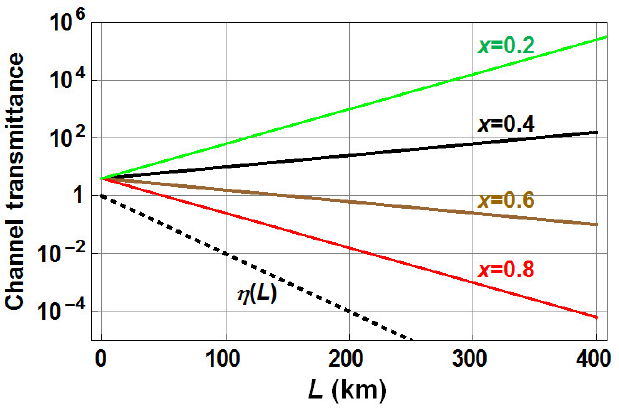}
\caption{\label{fig:Channel_trans}
The channel transmittance as a function of distance. 
The dashed line corresponds to the original channel 
without the tele-amplificcation. 
}
\end{center}
\end{figure}

\begin{figure}
\begin{center}
\includegraphics[width=0.92\linewidth]
{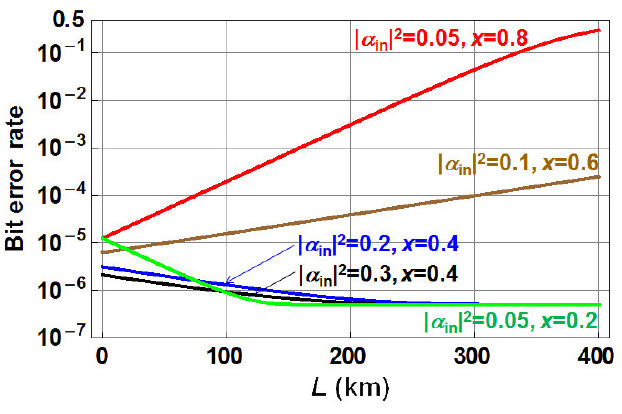}
\caption{\label{fig:BER}
The BER as a function of distance. 
}
\end{center}
\end{figure}

\begin{figure}
\begin{center}
\includegraphics[width=0.92\linewidth]
{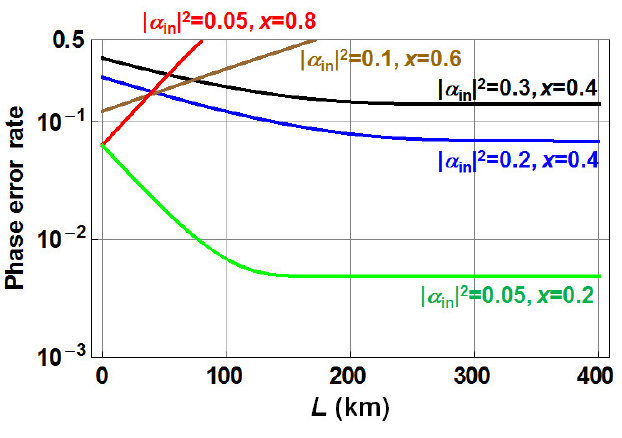}
\caption{\label{fig:PER}
The PER as a function of distance. 
}
\end{center}
\end{figure}

The essential effect brought by the tele-amplification 
is simply the improvement of the channel transmittance 
\beq
\eta(L) \rightarrow  g(x,L)^2 \eta(xL)
\eeq
but on the other hand also
the reduction of the detection rate 
due to the additional filtering at the relay node. 
We plot the channel transmittance in Fig. \ref{fig:Channel_trans}. 
For $x<0.5$, the channel turns to be an amplifier as a whole. 
It is clearly seen in the BER and PER as shown in 
Figs. \ref{fig:BER} and Fig. \ref{fig:PER}. 
They also decrease with the distance (black, blue and green lines). 
Actually there is no degradation of the signal-to-noise ratio 
as the distance increases. 
So the sudden fall of the key generation probability 
at a certain distance 
due to the dark count noise does not appear 
(see Fig. \ref{fig:Key_gen_prob}). 
For $x>0.5$, on the other hand, 
the BER and PER increase with the distance. 
In particular, the PER is the dominant error.


\begin{figure}[H]
\begin{center}
\includegraphics[width=0.92\linewidth]
{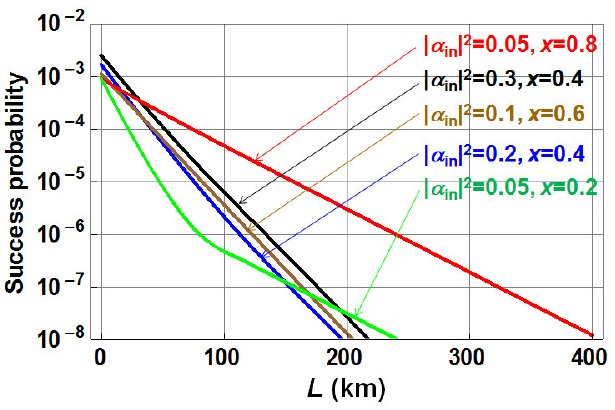}
\caption{\label{fig:Success_rate}
The success rate of the tele-amplification 
as a function of distance. 
}
\end{center}
\end{figure}

The success rate of the tele-amplification 
decreases with the distance as shown in 
Fig. \ref{fig:Success_rate}. 
This directly leads to 
the decrease of the key generation probability.

\end{document}